\documentclass[12pt,a4paper,final]{iopart}

\usepackage{iopams}  
\usepackage{graphicx}
\usepackage[breaklinks=true,colorlinks=true,linkcolor=blue,urlcolor=blue,citecolor=blue]{hyperref}

\newcommand{\be}{\begin{eqnarray}}
\newcommand{\ee}{\end{eqnarray}}

\def\tQ{\tau_{\rm Q}}
\def\teq{t_{\rm eq}}

\def\l{\lambda}

\begin{document}

\title[Quench dynamics of the extended Bose-Hubbard model]
{Dynamics of first-order quantum phase transitions in extended
Bose-Hubbard model: \\
From density wave to superfluid and vice-versa}

\author{Keita Shimizu$^{1}$, Takahiro Hirano$^1$, Jonghoon Park$^1$, Yoshihito Kuno$^{2}$ and Ikuo Ichinose$^{1}$}
\address{$^1$ Department of Applied Physics, Nagoya Institute of Technology, Nagoya 466-8555, Japan}
\address{$^2$ Department of Physics, Graduate School of Science, Kyoto University, Kyoto 606-8502, Japan}
\ead{k.shimizu.268@nitech.jp}



\begin{abstract}
In this paper, we study the nonequilibrium dynamics of the Bose-Hubbard model
with the nearest-neighbor repulsion by using
time-dependent Gutzwiller (GW) methods.
In particular, we vary the hopping parameters in the Hamiltonian
as a function of time, and investigate the dynamics of the system from
the density wave (DW) to the superfluid (SF) crossing a first-order phase transition
and vice-versa.
From the DW to SF, we find scaling laws for the correlation length and 
vortex density with respect to the quench time.
This is a reminiscence of the Kibble-Zurek scaling for continuous phase transitions
and contradicts the common expectation.
We give a possible explanation for this observation.
On the other hand from the SF to DW, the system evolution depends on 
the initial SF state.
When the initial state is the ground-state obtained by the static GW methods,
a coexisting state of the SF and DW domains
forms after passing through the critical point.
Coherence of the SF order parameter is lost as the system evolves.
This is a phenomenon similar to the glass transition in classical systems.
When the state starts from the SF with small local phase fluctuations, the system
obtains a large-size DW-domain structure with thin domain walls.
\end{abstract}

\pacs{67.85.Hj,	
03.75.Kk,	
05.30.Rt	
}
\vspace{2pc}
\noindent{\it Keywords}: Ultra-cold atomic gases, Bose-Hubburd model, Quantum dynamical phase transition, 
First-order phase transition, Superfluid, Density wave.
\maketitle

\section{Introduction}{\label{intro}}

In recent years, dynamics of quantum-many body systems is one of the 
most actively studied subjects in physics.
Process in which a system approach to an equilibrium is of fundamental
interests, and also evolution of system under a quench has attracted many
physicists. 
Nowadays, ultra-cold atomic gas systems play a very important role
for the study on these subjects because of their versatility, controllability
and observability~\cite{ob}.
Theoretical ideas proposed to understand transient phenomena are to be
tested by experiments on ultra-cold atomic systems.
This is one of examples of so-called quantum simulations
\cite{Nori,Cirac,coldatom1,coldatom2}.

For the second-order thermal phase transition,
time-evolution of systems under a change in temperature
has been studied extensively so far.
From the view point of cosmology, Kibble \cite{kibble1,kibble2} claimed that
the phase transitions lead to disparate local choices of the broken
symmetry state and as a result, topological defects called cosmic strings are generated.
Later, Zurek \cite{zurek1,zurek2,zurek3} pointed out that a similar phenomenon 
is realized in laboratory experiments on
the condensed matter systems like the superfluid (SF) of $^4$He.
After the above seminal works, many theoretical and experimental studies on 
the Kibble-Zurek (KZ) mechanism have appeared~\cite{IJMPA}.
Concerning to experiments on Bose-condensed ultra-cold atomic gases, 
the correlation length of the SF and the rate of topological defect formation
were measured and the KZ scaling hypothesis was examined~\cite{navon,Chomaz}.

To study dynamics of quantum many-body systems,
the parameters in the Hamiltonian are varied through a quantum phase transition (QPT),
i.e., the quantum
quench~\cite{Chomaz,dziarmaga,pol,Zoller,Fischer1,Fischer2,Be1,Be2,
sondhi,Zu1,Sonner,francuz,Zu2,SKHI},  
and the system evolution is observed.
Experiments on this problem have been already done 
using the various ultra-cold atomic gases~\cite{Chen,Braun,Anquez,clark,cui,bernien}.
Among them,
works in Refs.~\cite{Chen,Braun} questioned the applicability of the KZ scaling
theory to the QPT, whereas Refs.~\cite{Anquez,clark} concluded that
the observed results were in good agreement with the KZ scaling law.

In this paper, we focus on the two-dimensional (2D) Bose-Hubbard model 
(BHM)~\cite{BHM1,BHM2}, 
which is a canonical model of the bosonic ultra-cold atomic gas systems
in an optical lattice.
In particular, we add nearest-neighbor (NN) repulsions between atoms. 
Then, the resultant system
is described by an extended Bose-Hubbard model (EBHM).
As a result, a parameter region corresponding to the density wave (DW) appears 
in the ground-state phase diagram, in addition to the Mott insulator and SF.
Near the half-filling, there exists a first-order phase transition between
the SF and DW \cite{firstO}.
We shall study the quench dynamics of the EBHM on passing across the SF and DW
phase boundary. 
There are only a few works for the dynamical properties of quantum systems
at first-order phase transitions under a quench \cite{firstPT1,firstPT2,firstPT3}, 
and therefore 
detailed study on that problem is desired.

This paper is organized as follows.
In Sec.~2, we introduce the EBHM and explain the Gutzwiller (GW) methods,
which are used in the present work.
In Sec.~3, quench dynamics of the first-order phase transition from 
the DW to SF is studied.
Behavior of SF and DW orders are investigated by solving the Schr\"{o}dinger
equation by means of time-dependent GW (tGW) methods.
We focus on the order parameters, correlation length, vortex number, etc,
in particular, scaling laws of these quantities with respect to the quench time 
$\tau_{\rm Q}$.
Contrary to the common expectation, we find that scaling laws hold
for the correlation length and vortex density. 
In Sec.~4, we give a possible explanation of the observed results
from viewpoint of the SF bubble-nucleation process.
We employ a time-dependent Ginzburg-Landau theory and show that  
scaling laws with small deviations from the KZ scaling hold in the vicinity
of a triple point in the phase diagram.
Applicability of the GW methods is also discussed there.
In Sec.~5, we study the time evolution of the system from the SF to DW
crossing the first-order phase transition.
We find that even for very slow quench, a genuine DW does not form if
we start the time evolution with the ground-state obtained by the static
GW methods.
Numerical result shows that a coexisting state of the SF and DW appears instead.
On the other hand, if SF states with small coherent phase fluctuations are
employed as an initial state, the system acquires a DW domain structure of large
size with thin domain walls.
Section 6 is devoted for conclusion.
In appendix, we show the results obtained for the hard-core Bose-Hubbard model,
in which the first-order phase transition between the DW and SF exists as in the 
soft-core system of the present work.
We discuss the behavior of the correlation length and vortex density compared to 
the soft-core case.

\section{Extended Bose-Hubbard model and slow quench}{\label{BHM}}

We consider the EBHM 
whose Hamiltonian is given by \cite{Dutta},
\begin{eqnarray}
H_{\rm BH}&=&-J\sum_{\langle i,j \rangle}(a^\dagger_i a_j+\mbox{H.c.})
+{U \over 2}\sum_in_i(n_i-1)  \nonumber\\
&&+V\sum_{\langle i,j \rangle}n_in_j-\mu\sum_in_i,
\label{HEBH}
\end{eqnarray}
where $\langle i, j\rangle$ denotes NN sites of a square lattice,
$a_i^\dagger \ (a_i)$ is the creation (annihilation) operator of boson at site $i$,
$n_i=a^\dagger_i a_i$, and $\mu$ is the chemical potential.
$J(>0)$ and $U(>0)$ are the hopping amplitude and the on-site repulsion, respectively.
We also add the NN repulsion with the coefficient $V$, which plays an important role
in the present work.

\begin{figure}[t]
\centering
\begin{center}
\includegraphics[width=8cm]{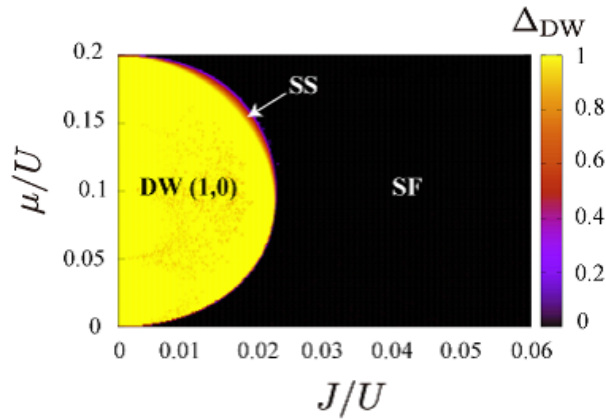}
\end{center}
\vspace{-0.5cm}
\caption{Ground-state phase diagram of the extended Bose-Hubbard model
for $V=0.05$ obtained by the static GW methods.
There exist three phases, the density wave (DW), superfluid (SF) and 
supersolid (SS).
Mean particle density $\rho\approx 1/2$.
}
\label{groundstate}
\end{figure}
\begin{figure}[t]
\centering
\begin{center}
\includegraphics[width=7cm]{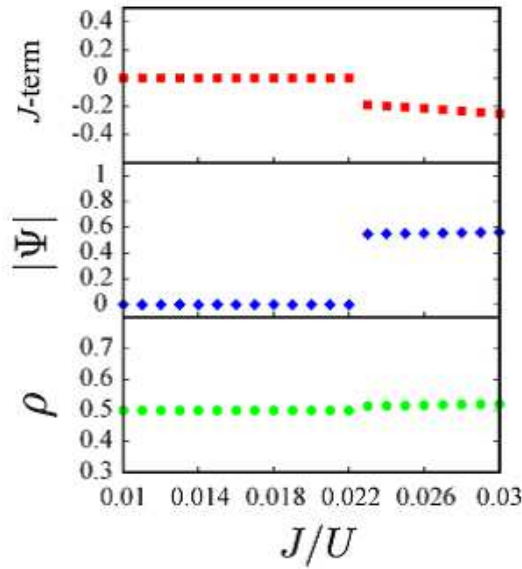}
\end{center}
\caption{Physical quantities in the DW and SF critical region in various system sizes;
the hopping $J$-term energy, amplitude of SF order ($|\Psi|$), 
and mean density ($\rho$).
The obtained results show that the phase transition is of first order as dictated
by Landau-Ginzburg-Wilson paradigm.
Critical point is estimated as $J_c/U\approx 0.022$.}
\label{static}
\end{figure}

In this study, we are interested in cases near the half filling, i.e.,
$\rho\equiv {1 \over N_s}\sum_i\langle n_i\rangle\approx 1/2$, 
where $N_s$ is the total number of the lattice sites, and 
we take $N_s=64\times 64$ or $100\times 100$ for the practical calculation.
We set $U=1$ as the energy unit, and time $t$ is measured in the unit $\hbar/U$.
We investigated the system in Eq.(\ref{HEBH}) by using 
the static GW approximation and show obtained ground-state phase diagram
in Fig.~\ref{groundstate} for $V/U=0.05$.
There exist three phases, i.e., the DW, SF and supersolid (SS) although the area of 
the SS in the phase diagram is small for $V/U=0.05$.
We also show the system energy, particle density and amplitude of the SF order
parameter,
$|\Psi|\equiv {1 \over N_s}\sum_i|\Psi_i|$, where $\Psi_i\equiv \langle a_i\rangle$,
in Fig.~\ref{static} for $\mu/U=0.1$.
From the results in Fig.~\ref{static}, it is obvious that the system exists near the 
half filling $\rho\approx 1/2$, and a first-order phase transition between the DW and SF
takes place at $J_c/U\simeq 0.022$ as a finite jump in $|\Psi|$ indicates.
The existence of the first-order phase transition is quite plausible as the DW and SF
have both the own long-range order.
In recent paper~\cite{KZIII}, we studied the EBHM for $V/U=0.375$ and near 
the unit filling $\rho\approx 1$.
There exists a substantially finite region of the SS in addition to the DW and SF.
These three phases are separated by two second-order phase transitions.
This result is in agreement with the quantum Monte-Carlo study~\cite{QMC}.

\begin{figure}[t]
\centering
\begin{center}
\includegraphics[width=9cm]{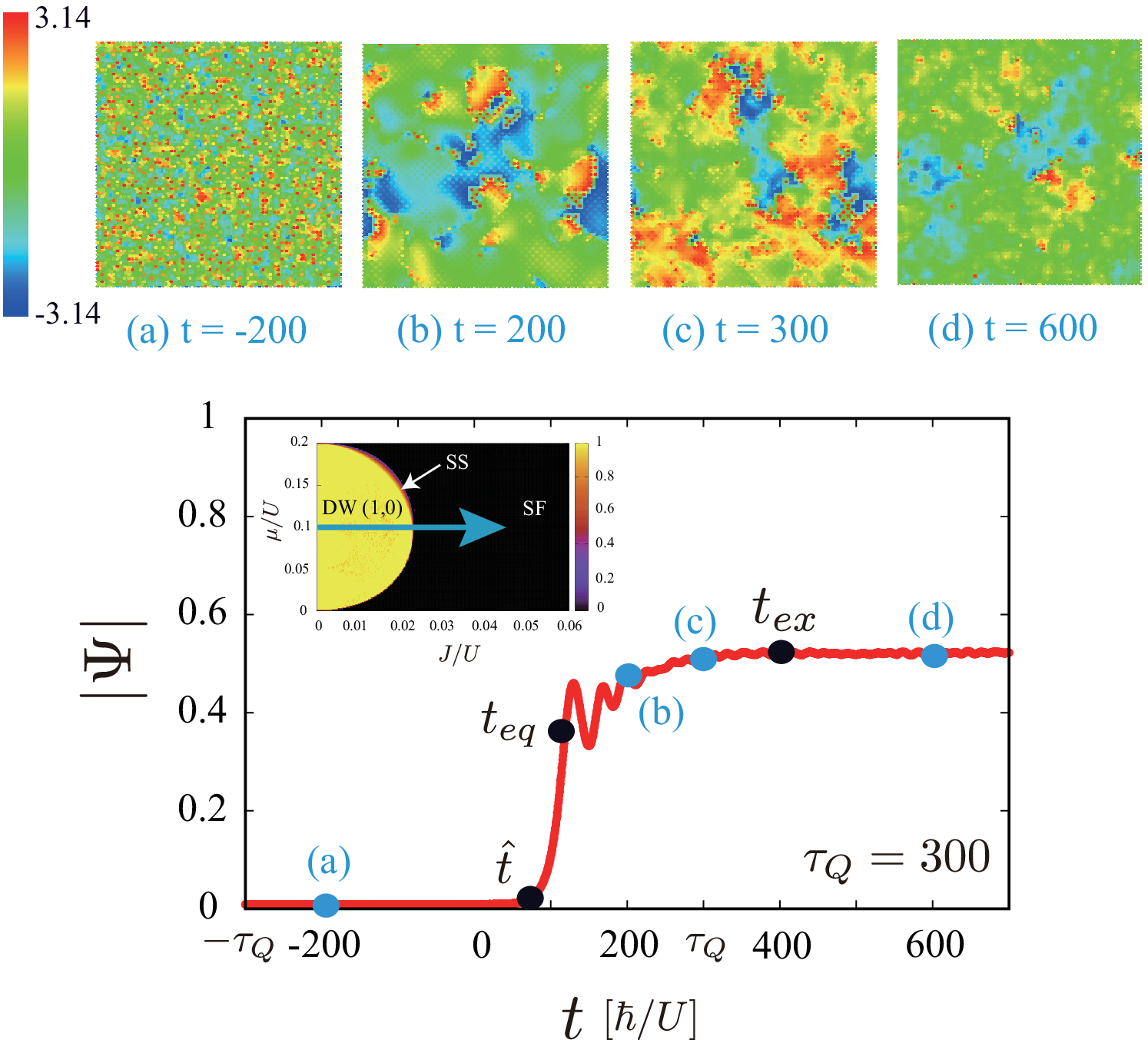}
\includegraphics[width=7cm]{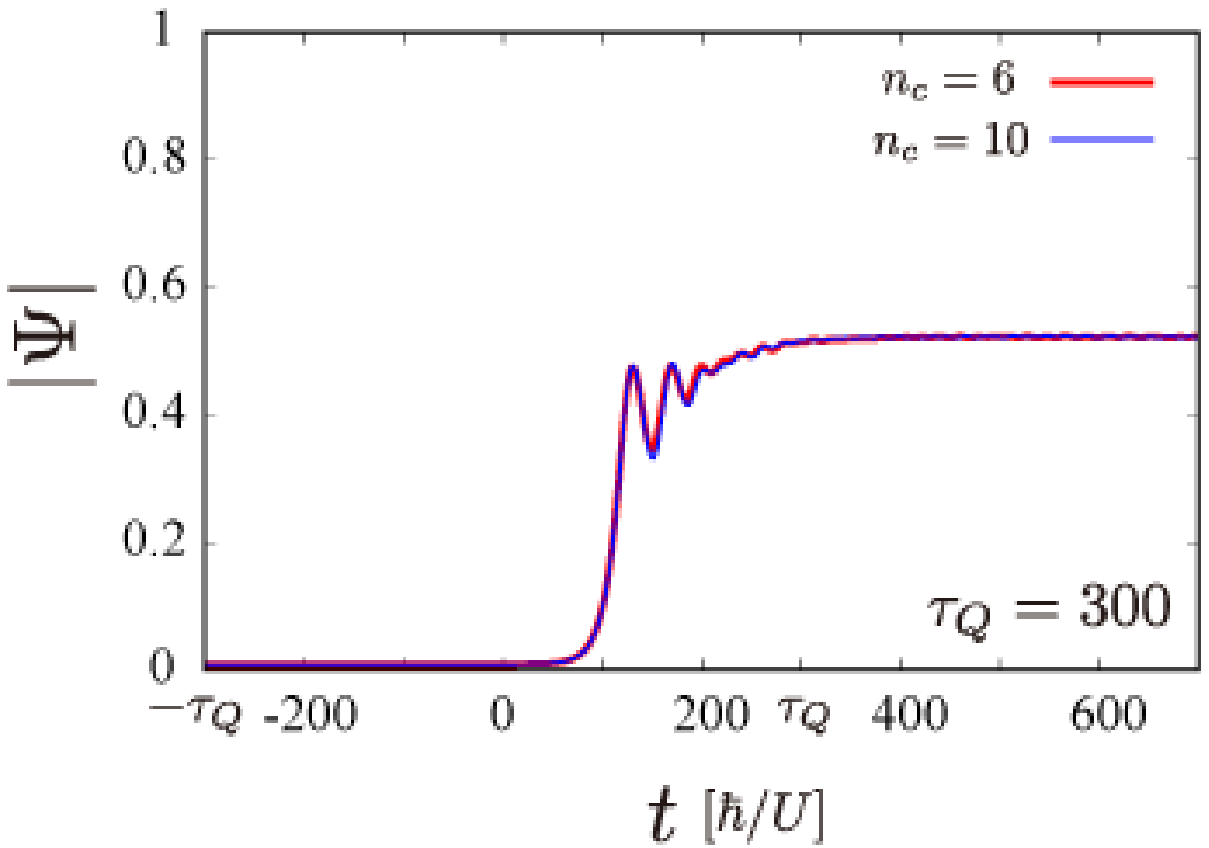}
\end{center}
\caption{(Upper panel) Phase of the SF order parameter $\Psi_i$
for $\tau_{\rm Q}=300$ as a function of time.
(Middle panel) Amplitude of the SF order parameter $\Psi_i$
for $\tau_{\rm Q}=300$ as a function of time.
Relevant times $\hat{t}$ and $t_{\rm eq}$ are $\hat{t}\approx 70$ and 
$t_{\rm eq}\approx 120$, respectively.
On the other hand, $t_{\rm ex}\approx 400$, at which the oscillation of
 $|\Psi|$ terminates.
From $t_{\rm eq}$ to $t_{\rm ex}$, coarsening process of the phase of $\Psi_i$ 
takes place in large scales \cite{SKHI}.
(Lower panel) Calculation of $|\Psi|$ in the $n_c=10$ case is also shown.
It is in good agreement with that of $n_c=6$.
}
\label{SForder}
\end{figure}

In the following, we shall study dynamics of the system under ``slow quenchs".
To this end, we employ the tGW 
methods~\cite{tGW1,tGW2,tGW3,tGW4,tGW5,tGW6,aoki}. 
In the tGW approximation, the Hamiltonian of the EBHM in Eq.(\ref{HEBH})
is approximated by a single-site Hamiltonian $H_i$, which is derived
by introducing the expectation value $\Psi_i=\langle a_i \rangle$,
\begin{eqnarray}
&&H_{\rm GW}=\sum_i H_i,   \nonumber\\
&&H_i=-J\sum_{j\in i{\rm NN}}(a^\dagger_i\Psi_j+\mbox{H.c.})
+{U \over 2}n_i(n_i-1)  \nonumber\\
&&\hspace{1cm}+V\sum_{j\in i{\rm NN}}n_i\langle n_j\rangle-\mu n_i,
\label{HGW}
\end{eqnarray}
where $i{\rm NN}$ denotes the NN sites of site $i$, and Hartree-Fock type 
approximation has been used for the hopping and NN repulsion.
To solve the quantum system $H_{\rm GW}$ in Eq.(\ref{HGW}),
we introduce GW wave function,
\begin{eqnarray}
|\Phi_{\rm GW}\rangle=\prod^{N_s}_i\Big(\sum^{n_c}_{n=0}
f^i_n(t)|n\rangle_i\Big),
\;\; \hat{n}_i|n\rangle_i=n|n\rangle_i,
\label{GW}
\end{eqnarray}
where $n_c$ is the maximum number of particle at each site, and we mostly take
$n_c=6$ in the present work.
Some quantities are calculated with $n_c=10$ to verify that $n_c=6$ is large enough
for the study of the half filling case.
See Fig.~\ref{SForder} and Fig.~\ref{correlationtime}.
In terms of $\{f^i_n(t)\}$, the order parameter of the SF is given as,
\begin{eqnarray}
\Psi_i=\langle a_i \rangle=\sum^{n_c}_{n=1}\sqrt{n}f^{i\ast}_{n-1}f^i_n,
\label{BEC}
\end{eqnarray}
and $\{f^i_n(t)\}$ are determined by solving the following Schr\"{o}dinger equation
for various initial states,
\begin{eqnarray} i\hbar \partial_t |\Phi_{\rm GW}\rangle
=H_{\rm GW}(t)|\Phi_{\rm GW}\rangle.
\label{SEq}
\end{eqnarray}
The time dependence of $H_{\rm GW}(t)$ in Eq.(\ref{SEq}) comes from the quench
$J\to J(t)$ with fixed $U$ and $V$ as explained in the following section.
Practically, the time evolution above is calculated by the 
fourth-order Runge-Kutta method.


\section{Dynamics of phase transition from density wave to superfluid}\label{DWtoSF}

We first study the dynamics from the DW to SF.
In this section, the hopping amplitude is varied as 
\begin{eqnarray}
{J(t)-J_c \over J_c}\equiv \epsilon(t)={t \over \tau_{\rm Q}},
\label{protocol}
\end{eqnarray}
where $\tau_{\rm Q}$ is the quench time, which is a controllable parameter in
experiments.
We employed 10 samples as the initial state at $t=-\tau_{\rm Q}$
(i.e., $J(-\tQ)=0$), which have
the DW order with small local density fluctuations from the perfect DW.
Then, we solve Eq.(\ref{SEq}) to obtain $|\Phi_{\rm GW}\rangle$.
Physical quantities for which scaling lows are examined are obtained by 
averaging over samples.
The linear quench in Eq.(\ref{protocol}) is terminated at $t=t_{\rm f}$ with
$J(t_{\rm f})=0.044(>J_c)$ in the numerical study.
Subsequent behavior of the system is also observed to see how the system
approaches to an equilibrium.

\begin{figure}[h]
\centering
\begin{center}
\includegraphics[width=7cm]{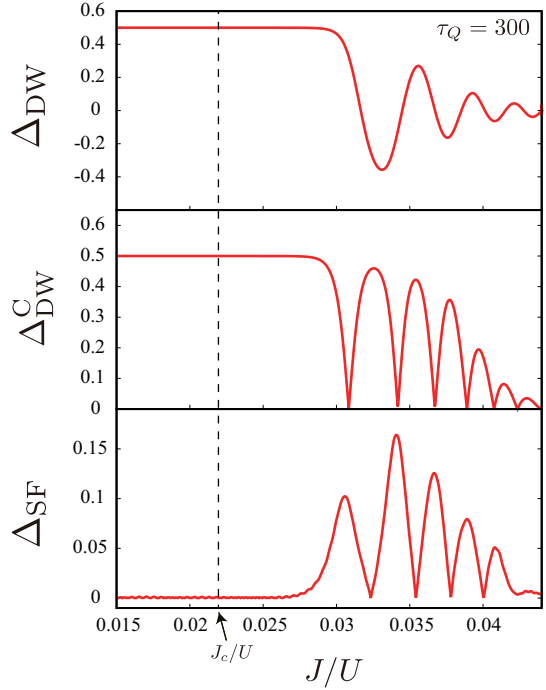}
\end{center}
\caption{$\Delta_{\rm DW}$,  $\Delta^{\rm c}_{\rm DW}$
and $\Delta_{\rm SF}$ as a function of time
for $\tau_{\rm Q}=300$.
After passing the equilibrium critical point $J_c/U\simeq 0.022$, the both quantities
start to evolve with oscillations.
}
\label{Diff}
\end{figure}

We show the typical behavior of $|\Psi|$ as a function of $t$ in Fig.~\ref{SForder} 
for $\tau_{\rm Q}=300$.
At $t=0$, the system crosses the critical point at $J_c/U\simeq 0.022$.
After crossing the critical point, $|\Psi|$ remains vanishingly small for some period,
and then it develops very rapidly.
After the rapid increase, $|\Psi|$ starts to fluctuate and coarsening of 
the phase of the SF order parameter takes place there~\cite{SKHI}.
$\hat{t}$ in Fig.~\ref{SForder} is defined as $|\Psi(\hat{t})|=2|\Psi(0)|$,
and $t_{\rm eq}$ is the time at which the oscillation of $|\Psi|$ starts.
Similarly, $t_{\rm ex}$ is the time at which that oscillation terminates.

Similar behavior to the above was observed in the Mott to SF quench dynamics and 
examined carefully~\cite{SKHI}.
Compared with the Mott to SF dynamics, the SF amplitude $|\Psi|$ is smaller,
e.g., for $t>t_{\rm eq}$, $|\Psi|\sim (0.8-0.9)$ in the Mott to SF transition, whereas
$|\Psi|\sim 0.5$ in the present case. 
This difference simply comes from the difference of the mean particle density, i.e., 
$\rho \sim 1$ in the Mott to SF transition case.

The DW order parameters 
$\Delta_{\rm DW}\equiv {1 \over N_s}\sum_i(-)^i\langle n_i\rangle$,
$\Delta^{\rm C}_{\rm DW}\equiv {1 \over 2N_s}\sum_{\langle i,j\rangle}
|\langle (n_i-n_j)\rangle|$,
and the even-odd deference of the SF order parameter defined as
$\Delta_{\rm SF}\equiv {1 \over 2N_s}\sum_{\langle i,j\rangle}
||\Psi_i|-|\Psi_j||$
are shown in Fig.~\ref{Diff}.
These quantities exhibit fluctuations as a function of time until $J\approx 0.045$. 
These fluctuations are getting smaller, i.e., the system is approaching to a
homogeneous SF.
The system with other vales of $\tau_{\rm Q}$ exhibits a similar behavior, although
the reaction of the system starts at larger value of $J/U$ for
smaller value of the quench time $\tau_{\rm Q}$.

It is interesting to study the correlation length $\xi$ of the SF order parameter
and the vortex density $N_{\rm v}$ as a function of the quench time $\tau_{\rm Q}$.
These quantities are defined as follows;
\begin{eqnarray}
&&\langle \Psi_i^\ast \Psi_j\rangle \propto \exp (-|i-j|/\xi), \nonumber \\
&&N_{\rm v}=\sum_i|\Omega_i|, \nonumber \\
&&\Omega_i={1 \over 4}\Big[\sin (\theta_{i+\hat{x}}-\theta_i)
+\sin (\theta_{i+\hat{x}+\hat{y}}-\theta_{i+\hat{x}})
 \nonumber \\
&&\hspace{1cm} -\sin (\theta_{i+\hat{x}+\hat{y}}-\theta_{i+\hat{y}})
-\sin (\theta_{i+\hat{y}}-\theta_{i})\Big],
\end{eqnarray}
where $\theta_i$ is the phase of $\Psi_i$ ($\Psi_i=|\Psi_i|e^{i\theta_i}$)
and $\hat{x} \ (\hat{y})$ is the unit vector in the $x \ (y)$ direction.
For continuous second-order phase transitions, the KZ hypothesis 
predicts a scaling law such as $\xi\propto \tau_{\rm Q}^b$ and 
$N_{\rm v}\propto \tau_{\rm Q}^{-d}$.
Recently, applicability of the above KZ scaling law for 
{\em second-order quantum
phase transition} has been discussed for several quantum systems.
On the other hand for first-order phase transitions, it is commonly expected that
such a scaling law does not hold as the relaxation time  
cannot be defined properly.
For a classical statistical model, another type of scaling law was proposed
for first-order phase transitions~\cite{firstPT1}.
It should be also noted that off-equilibrium dynamics of a quantum Ising ring
was investigated recently and finite-size scaling laws for first-order phase 
transitions were proposed~\cite{firstPT11}.
There, off-equilibrium scaling variables were given in terms of an energy gap 
and quench time, and physical quantities were obtained as a function of time.

\begin{figure}[t]
\centering
\begin{center}
\includegraphics[width=7.5cm]{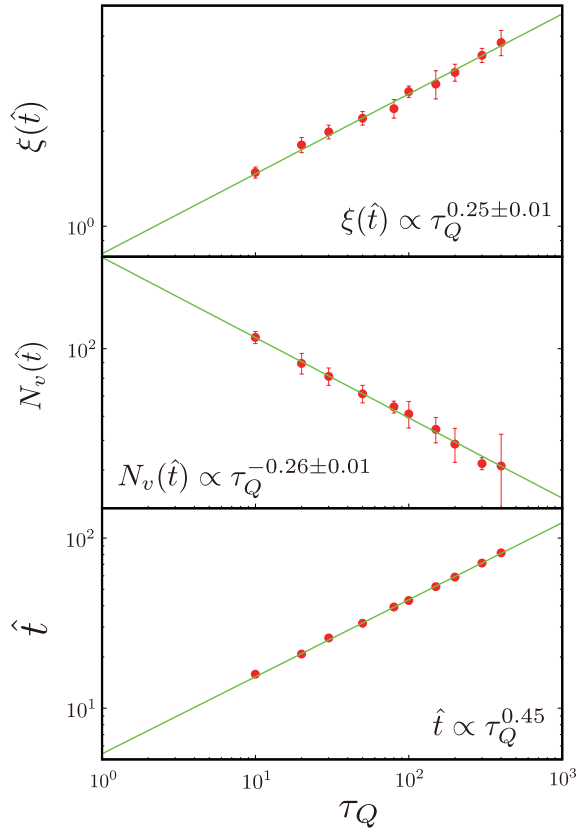}
\end{center}
\caption{Scaling laws observed for the correlation length $\xi$, vortex number 
$N_{\rm v}$ at $t=\hat{t}$, and $\hat{t}$
with respect to $\tau_{\rm Q}$.
}
\label{scalinghatt}
\end{figure}
\begin{figure}[t]
\centering
\begin{center}
\includegraphics[width=7.5cm]{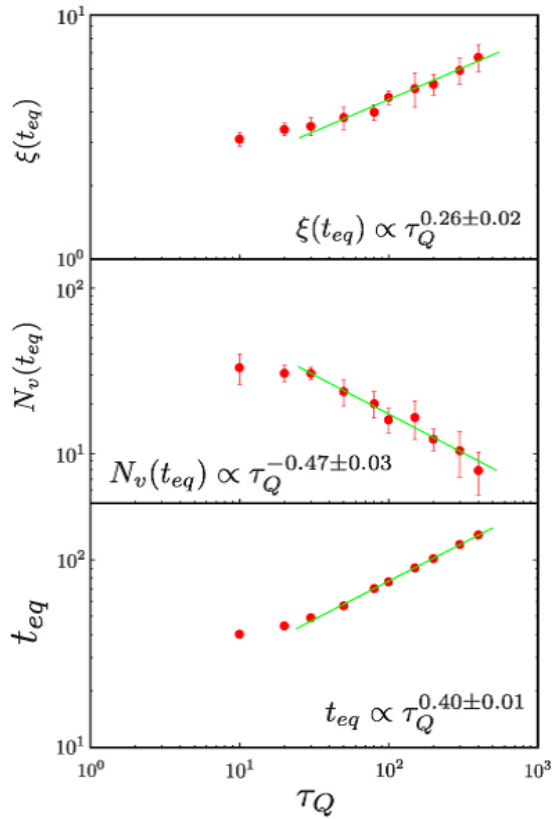}
\end{center}
\caption{Scaling lows observed for the correlation length, vortex number at
$t=\teq$, and $\teq$ with respect to $\tQ$.
}
\label{scalingteq}
\end{figure}

To see if scaling law exists or not, we measured $\xi$ and $N_{\rm v}$ at
$t=\hat{t}$ and $t=\teq$.
In the original KZ hypothesis for continuous phase transitions~\cite{IJMPA}, 
$\hat{t}$ is the time at which
the system re-enters an equilibrium after the freezing (or impulse) period.
On the other hand, $\teq$ is the time at which a coarsening process
of the SF phase coherence starts~\cite{SKHI}.

We show the obtained results in Figs.~\ref{scalinghatt} and \ref{scalingteq}.
The results show that
at $t=\hat{t}$, both $\xi$ and $N_{\rm v}$ satisfy the scaling law with exponents
$b=0.25$ and $d=0.26$, respectively, and also $\hat{t}\propto \tQ^{0.45}$.
On the other hand at $t=\teq$, data at each $\tQ$ exhibits slightly large fluctuations
but scaling laws for the correlation length, $N_{\rm v}$ and $\teq$ seem to exist
for $\tQ>20$.
The above results indicate that besides the KZ mechanism, there exists 
another mechanism to generate the scaling laws.
Possible explanation is given in Sec.~4.

It should be noted that
after passing the critical point, $\Delta_{\rm DW}$ and $\Delta_{\rm SF}$
have even-odd site fluctuations, and therefore, the system is not homogeneous.
We think that because of this inhomogeneity, the critical exponents of $\xi$ and 
$N_{\rm v}$ at $t=\hat{t}$ do not satisfy the expected relation such as $b=d/2$.
On the other hand at $t=\teq$, the system is rather homogeneous, and therefore
$b\sim d/2$.

In appendix, we consider the hard-core version of the EBHM and show 
the calculations of the scaling laws with respect to $\tQ$ in Fig.~\ref{HCtQ}.
There, $\xi(\hat{t})$ and $N_{\rm v}(\hat{t})$ fluctuate rather strongly.
This behavior comes from the fact that fluctuations of the particle number 
at each site is smaller compared with the soft-core case, and as a result,
the stability of the phase degrees of freedom of the SF order parameter
is weakened.

\begin{figure}[h]
\centering
\begin{center}
\includegraphics[width=6.5cm]{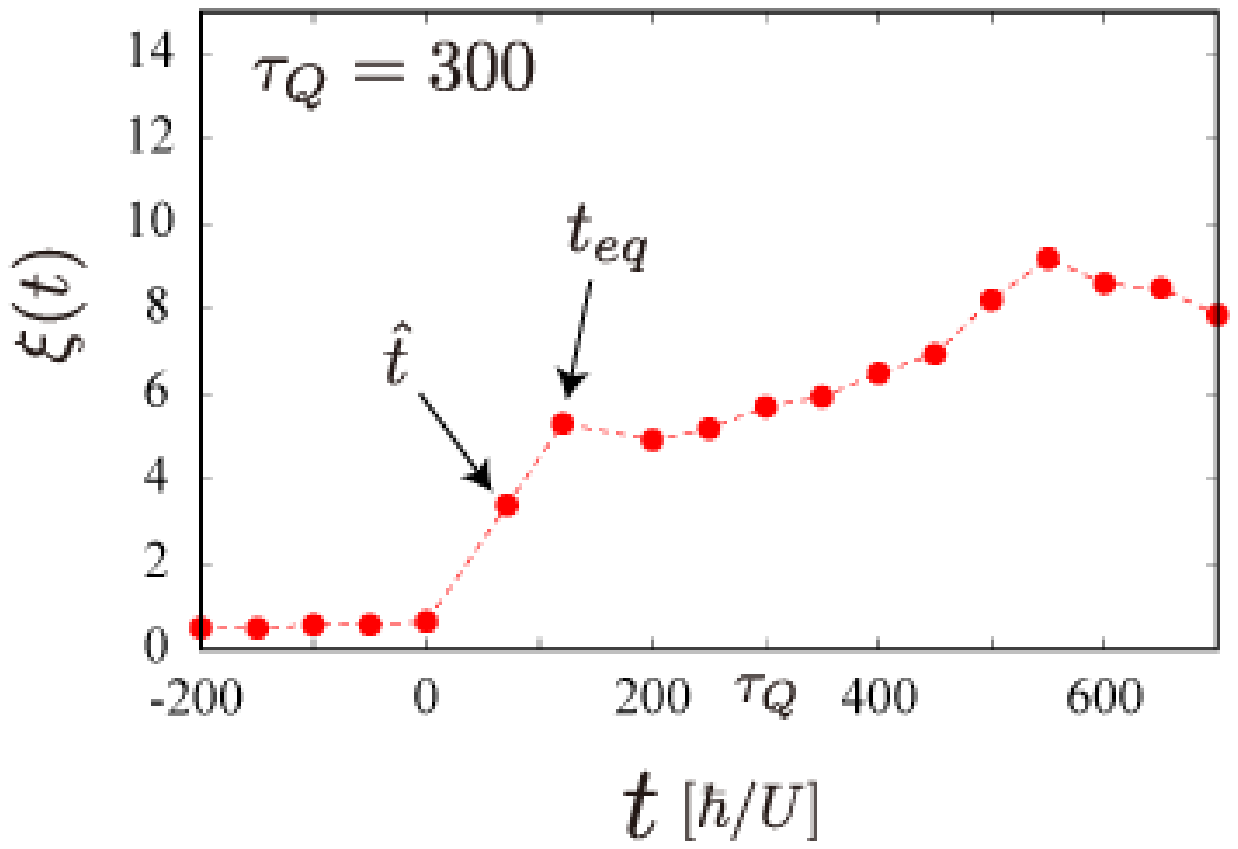}
\includegraphics[width=6.5cm]{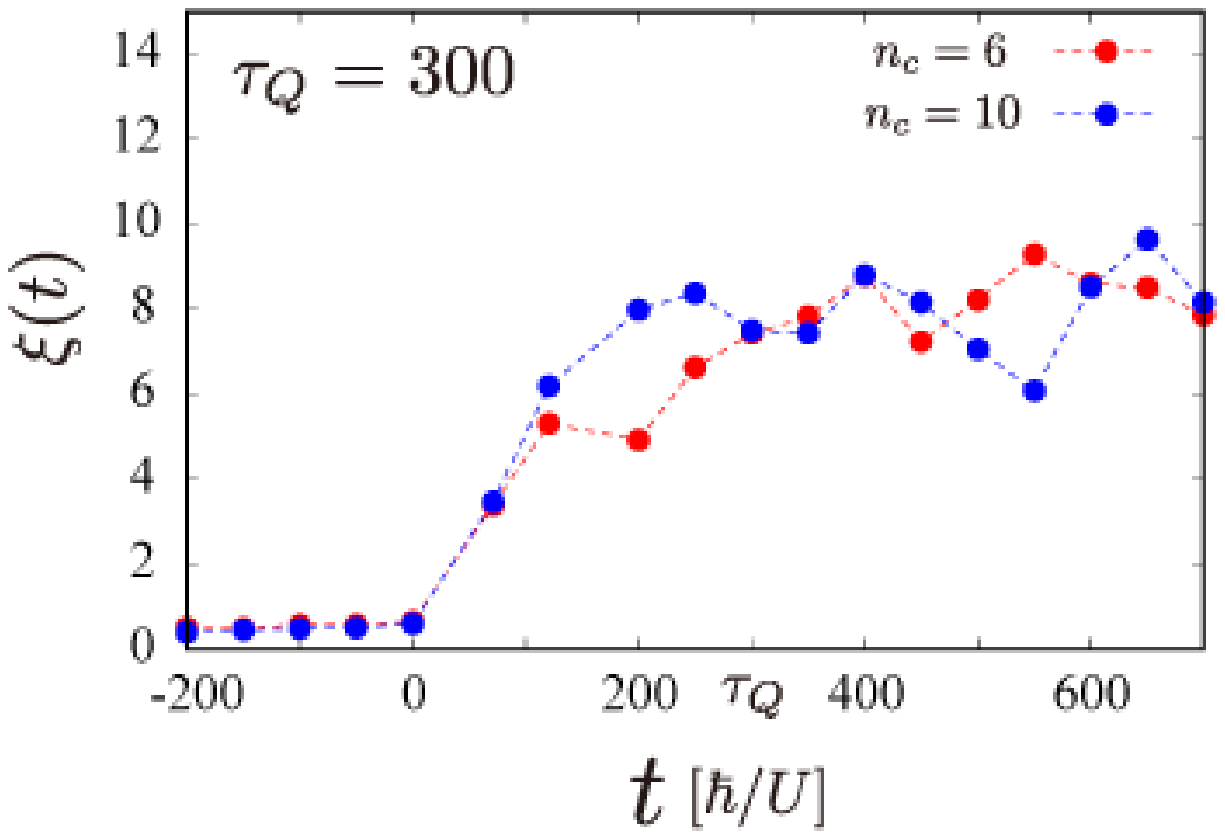} \vspace{0.5cm} \\
\includegraphics[width=5cm]{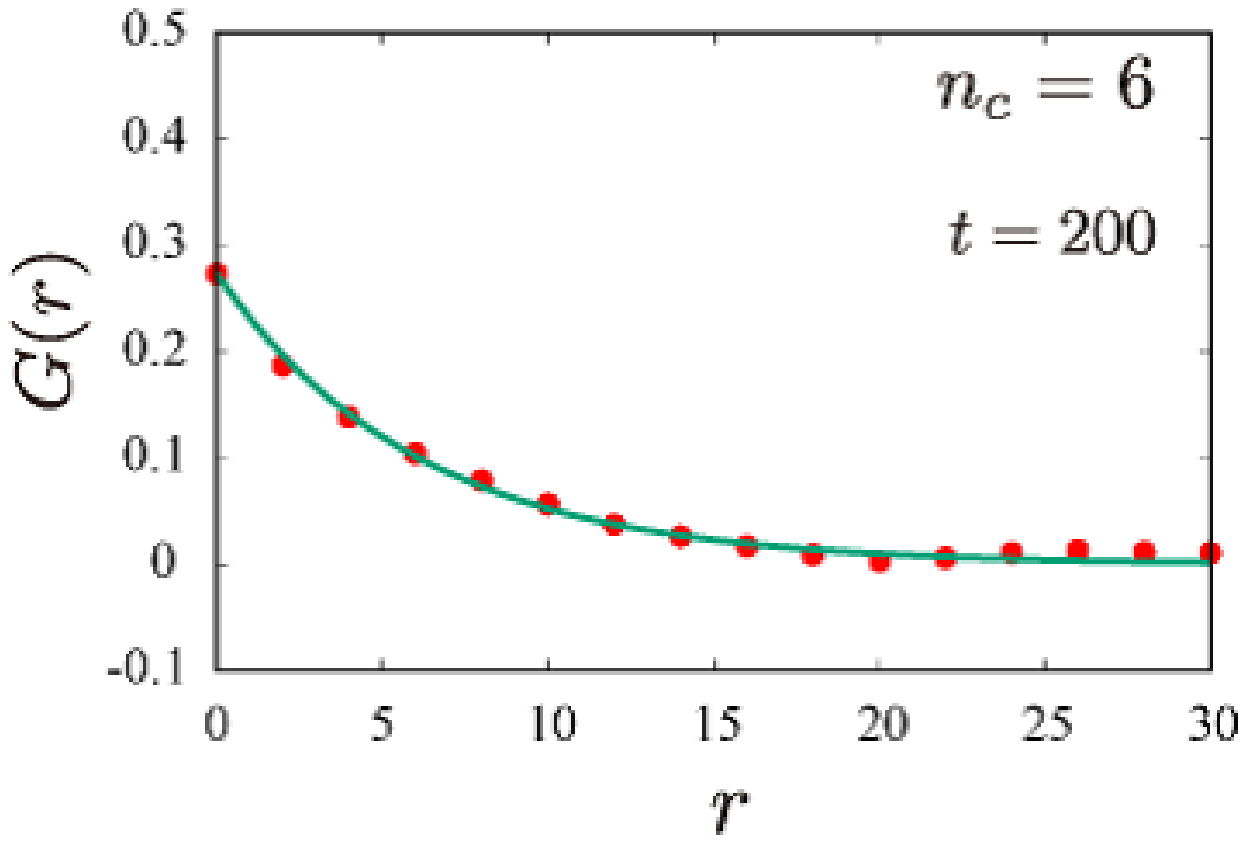}
\includegraphics[width=5cm]{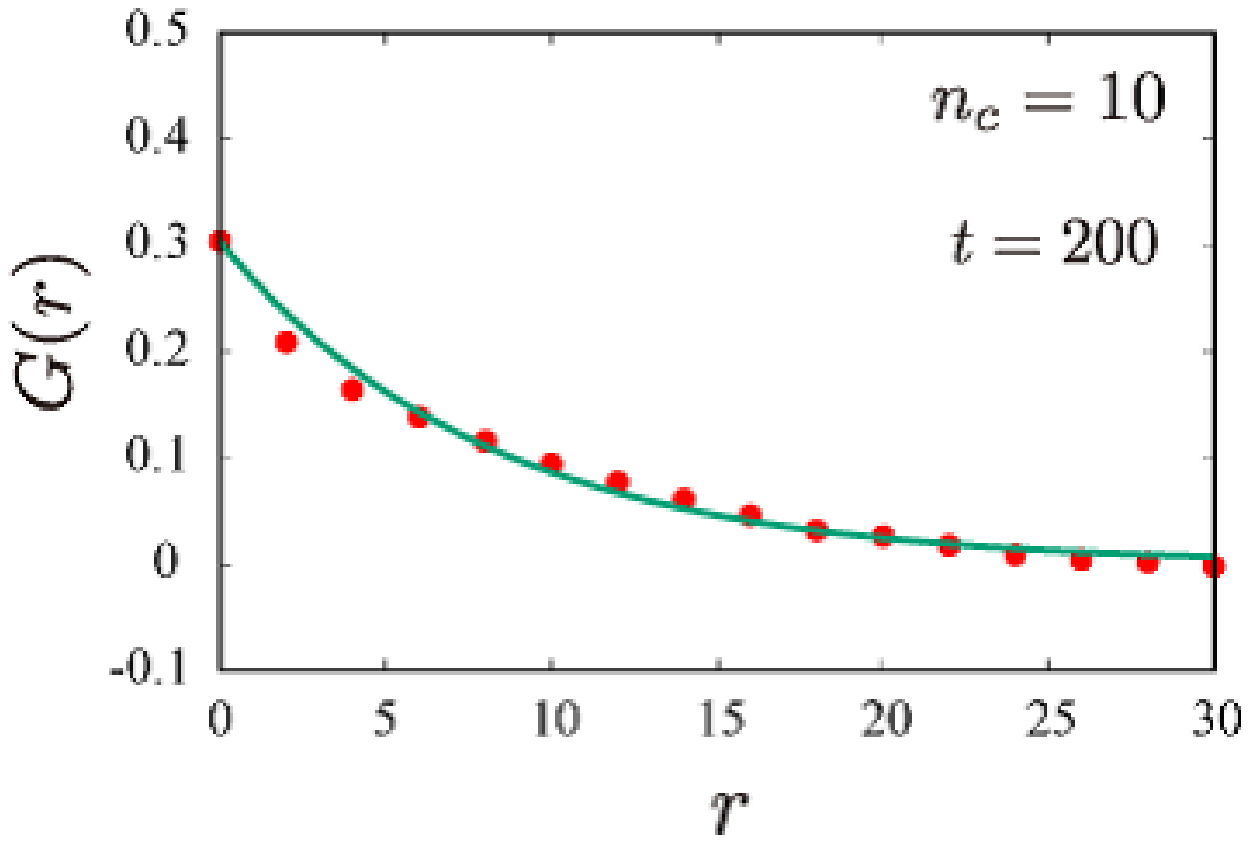} \\
\includegraphics[width=5cm]{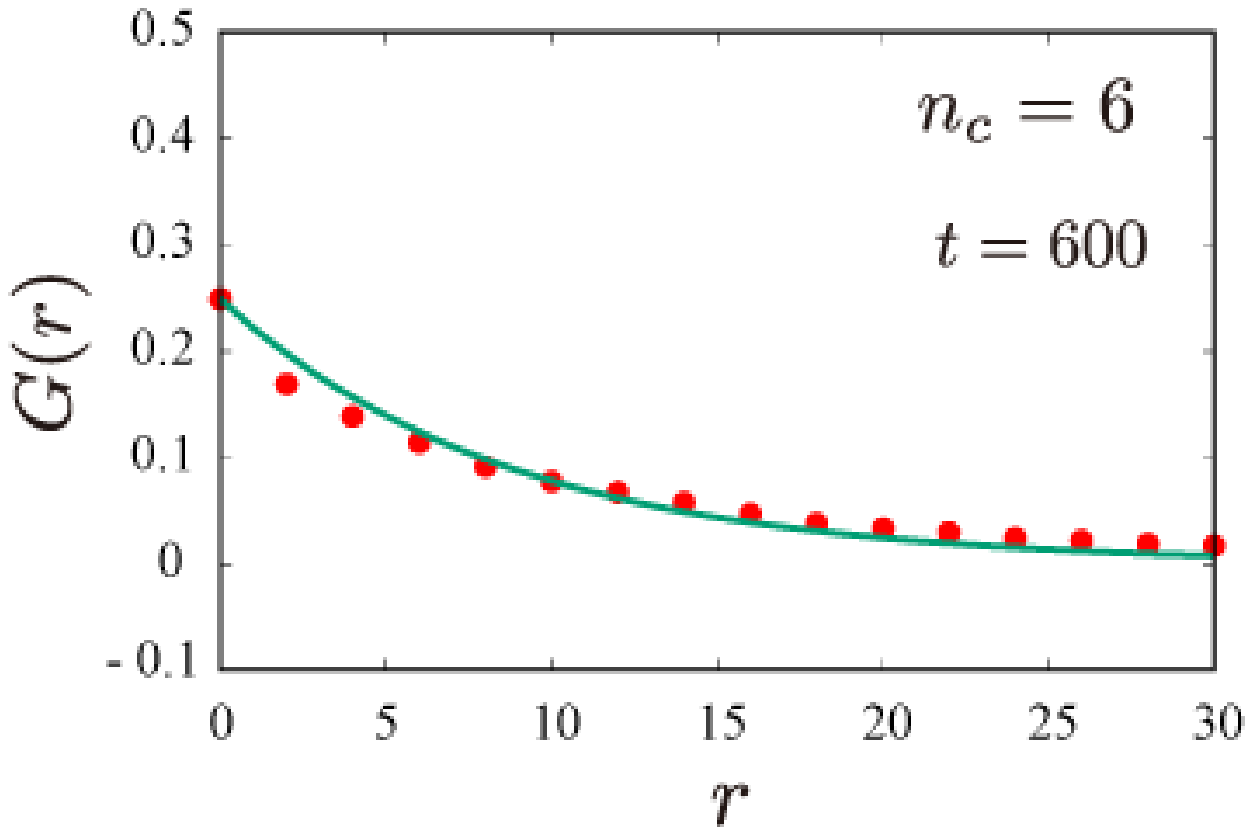}
\includegraphics[width=5cm]{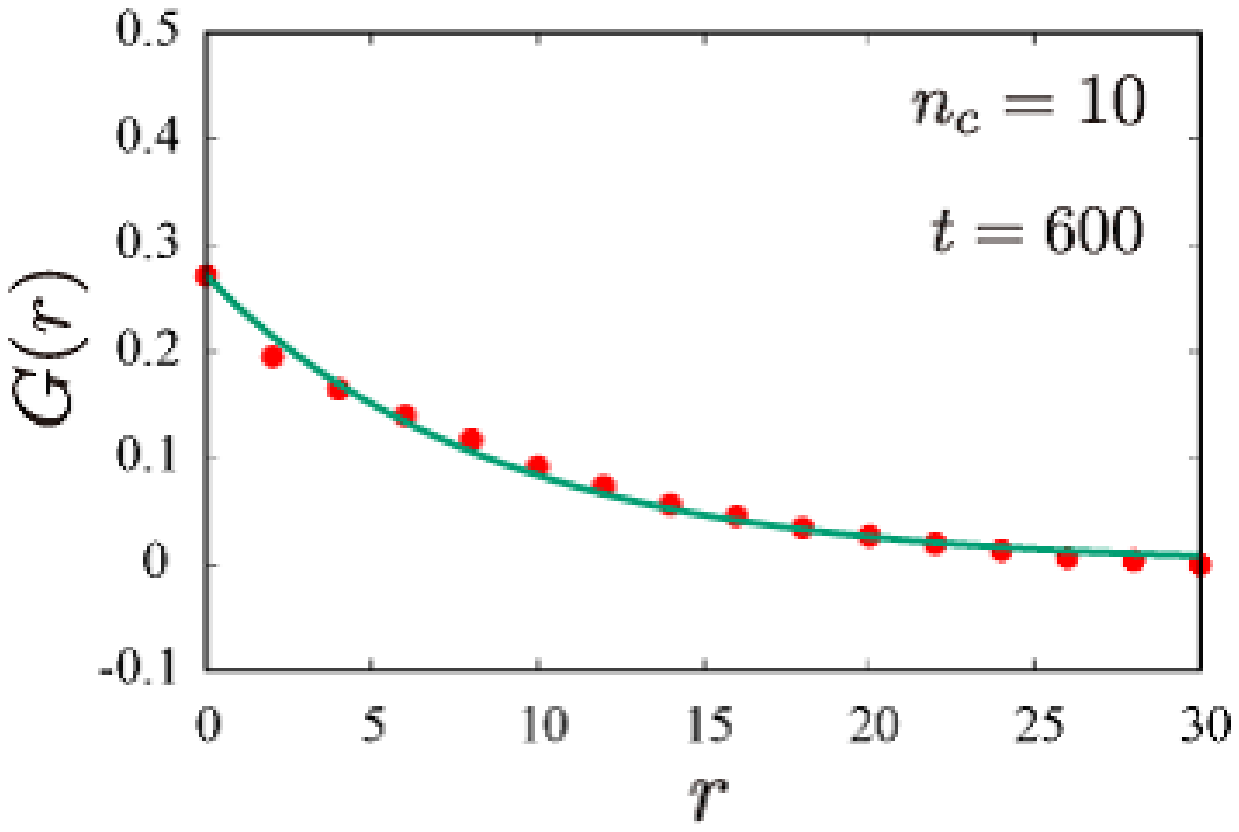}
\end{center}
\caption{(Upper left panel) For a typical initial state at $t=-\tQ$,
the correlation length is calculated as a function of time.
After passing $t=\teq$, increase of the correlation length becomes weak.
(Upper right panel) We also show the results for the $n_c=10$ case.
(Lower panels) The correlation functions 
$G(r)={1 \over 2N_s}\sum_i\langle a_i^\dagger a_{i+r}\rangle$ exhibit
very close behavior in the $n_c=6$ and $n_c=10$ cases.
}
\label{correlationtime}
\end{figure}

We terminate the linear quench at $t_{\rm f}=\tQ=300$.
After $t_{\rm f}$, the system approaches to an equilibrium as the results
in Figs.~\ref{SForder} and \ref{Diff} indicate.
It is interesting to see how the correlation length of the SF develops.
As the results in Fig.~\ref{correlationtime} show, 
the correlation length increases after passing the critical point as it is expected.
However, its increase gets weak at $t\sim\teq$, and it 
saturates at $t\sim 500$ and keeps a finite value.
To study the resultant phase, we measured $N_{\rm v}$ and found 
that there exist no vortices at $t>500$.
One may expect that the system settles in a {\em finite-temperature} ($T$) SF phase
for sufficiently large $t$ with an effective $T$, $T_{\rm eff}$.
The finite-$T$ SF in 2D has a quasi-long range order and the correlation length
diverges, i.e., the Kosterlitz-Thouless (KT) phase.
The above result seems to indicate that some other state is realized in the final
stage of the present process.
However, the system behavior may strongly depend on the average particle 
density $\rho$.
Further study is needed to clarify this interesting problem.
In fact, we studied this problem in the case of the mean particle density 
$\rho \approx 1$ and $V/U=0.375$~\cite{KZIII}.
In the quench process such as the DW $\to$ SS $\to$ SF, the correlation length 
continues to increase even for large $t$.
This result seems to indicate that a KT phase of the SF is realized there.


\section{Consideration by the Ginzburg-Landau theory}

In the previous section, we showed that the results obtained by the GW methods
indicate the scaling laws of $\hat{t}$, $\teq$ and the correlation length with
respect to the quench time $\tQ$.
It is interesting and also important to study the origin of these observations
from more universal and intuitive point of view.
To this end, the Ginzburg-Landau (GL) theory is quite useful.
In fact very recently, it was pointed out that the GL theory can drive the
scaling laws for the second-order phase transition by analytical transformation of 
the associated equations of motion~\cite{Niko}.
In this section, we first review the above derivation of the scaling laws for the ordinary
second-order phase transition, and then give an intuitive picture of the scaling laws
by using a classical solution representing decay of the false vacuum. 
Then, we extend the methods to the present case involving the SF and DW
order parameters.
This consideration also gives an insight about the physical meaning and limitation of the
GW methods.

\subsection{Second-order phase transition}

Let us start with the stochastic GL equation for a complex order parameter
(condensate) $\phi(\vec{r}, t)$,
\be
{\partial \phi \over \partial t}=\nabla^2_r \phi-{\epsilon(t) \over 2}\phi 
-{1\over 2}|\phi|^2\phi+\Theta(\vec{r},t),
\label{GLeq}
\ee
where $\Theta(\vec{r},t)$ represents the uncorrelated white-noise variables with
$\langle \Theta(\vec{r},t)\Theta(\vec{r}',t')\rangle=T \delta(\vec{r}-\vec{r}')
\delta(t-t')$ and $T$ is the temperature of particles ensemble not participating
the Bose-Einstein condensate.
As in Ref.~\cite{Niko}, we consider the critical parameter $\epsilon(t)$ such as
\be
\epsilon(t)=-\Big|{t\over \tQ}\Big|^\lambda \mbox{sgn} (t),
\label{epsilont}
\ee
where $\lambda$ is a parameter for the quench protocol.
Then, let us change variables as follows,
\be
\eta=\alpha t, \;\; \vec{\ell}=(\alpha)^{1/2}\vec{r}, \;\; 
\tilde{\phi}=\phi/(\alpha)^{1/2},
\label{change}
\ee
where $\alpha=\tQ^{-\l/(\l+1)}$.
In terms of the new variables, the equation of motion (\ref{GLeq}) leads to
\be
{\partial \tilde{\phi} \over \partial \eta}=\nabla^2_\ell \tilde{\phi}
-{1 \over 2}|\eta|^\l \mbox{sgn}(\eta)\tilde{\phi}
-{1 \over 2}|\tilde{\phi}|^2\tilde{\phi}+{1 \over \alpha}\Theta(\vec{\ell}, \eta).
\label{GLeq1}
\ee
In Eq.(\ref{GLeq1}), the $\tQ$-dependence in Eq.(\ref{epsilont}) disappears except 
the last white-noise term.
From the above fact, it is concluded in Ref.~\cite{Niko} that 
the $\tQ$-dependence of $\hat{t}$ and $\xi(\hat{t})$ are expected to follow
the transformation in Eq.(\ref{change}), and they are given as follows 
for sufficiently low $T$,
\be
\hat{t} \propto \alpha^{-1}=\tQ^{\l/(\l+1)}, \;\;\;
\xi(\hat{t}) \propto \alpha^{-1/2}=\tQ^{\l/2(\l+1)}.
\label{scalGL}
\ee
For the linear quench $\l=1$, $\hat{t}\propto \tQ^{1/2}$ and 
$\xi(\hat{t})\propto \tQ^{1/4}$.
The above estimations agree with those of the KZ scaling with the mean-field
exponents such as $\nu=1/2$ and $z=2$.

As we show, the above scaling transformation gives an intuitive picture that derives
the KZ scaling law.
To this end, we put $\Theta(\vec{r},t)=0$ in Eq.(\ref{GLeq}) and consider 
a static potential such as $\epsilon(t)=-\epsilon_0<0$.
In this case, the static ground state is given as $\phi=\sqrt{\epsilon_0}$.
To study the {\em sudden quench dynamics}, we consider the decay of the false vacuum 
$\phi=0$ to the true ground state $\phi=\sqrt{\epsilon_0}$.
In 1D case, a classical solution representing the decay is obtained as 
follows~\cite{firstPT3},
\be
\phi(t,x)=\sqrt{\epsilon_0}
\Big[1+\exp \Big({\sqrt{\epsilon_0} \over 2}(x-v_0t)\Big)\Big]^{-1},
\label{solution1}
\ee
where $v_0={3\sqrt{\epsilon_0} \over 2}$, and $\phi(t,-\infty)=\sqrt{\epsilon_0}$
and $\phi(t,\infty)=0$.
The solution Eq.(\ref{solution1}) obviously represents the situation in which the true
vacuum $\phi=\sqrt{\epsilon_0}$ born in the false vacuum expands with the speed $v_0$.

Let us consider the ``slow" quench dynamics and study bubble nucleation-evolution
process in the SF formation.
We expect that this process corresponds to the numerical studies in the previous sections.
We have to find the solution to Eq.(\ref{GLeq}) that describes a single SF-bubble
evolution in the false vacuum $\phi=0$, but we cannot find an exact solution.
However,
the above solution in Eq.(\ref{solution1}) suggests that a spherically-symmetric
solution in higher dimensions and also for the time-dependent $\epsilon(t)$  
has the following form for $\epsilon(t)<0$,\footnote{Solution in Eq.(\ref{solution2})
might be regarded a solution in the slow quench limit, in which the time-derivative 
of $\epsilon(t)$ is small. However, it also satisfies the scaling transformation with
$\epsilon(t)$ in Eq.(\ref{epsilont}). See the discussion below.}
\be
\phi_s(\vec{r},t)=\sqrt{|\epsilon(t)|}F\Big(\sqrt{|\epsilon(t)}|(r-v_tt)\Big),
\;\;\; r>0,
\label{solution2}
\ee
where $v_t=C_0\sqrt{|\epsilon(t)|}$ with a certain constant $C_0$, and 
$F(x)$ is a decreasing function such as $F(-\infty)=1$ and $F(\infty)=0$.
In fact, we can show that the function $\phi_s(\vec{r},t)$ in 
Eq.(\ref{solution2}) satisfies the scaling transformation in Eq.(\ref{change})
for the time-dependent $\epsilon(t)$ in Eq.(\ref{epsilont}), i.e.,
\be 
\tilde{\phi}_s(\eta,\vec{\ell})=\phi_s/(\alpha)^{1/2}
=\sqrt{\eta^\l}F\Big(\sqrt{\eta^\l}(\ell-v(\eta)\eta)\Big),
\;\; v(\eta)=C_0\sqrt{\eta^\l},
\label{solution3}
\ee
does not depend on $\tQ$.
As far as the above picture holds in the time evolution of the system,
Eq.(\ref{solution3}) implies that typical events and phenomena are 
observed similarly in systems with various $\tQ$'s, and corresponding times have 
$\tQ$-dependence such as $\tQ^{\l/(\l+1)}$.
For example, we numerically obtained $\hat{t}$ and $\teq$ for various $\tQ$'s in 
Sec.~3 by starting with qualitatively the same initial states.
These values are related to $\tQ$-independent $\hat{\eta}$
and $\eta_{\rm eq}$ that are obtained by the rescaled picture from 
Eq.(\ref{solution3}), i.e., 
$\hat{t}$ and $\teq$ in the $\tQ$-system are given by
$\hat{t}=\tQ^{\l/(\l+1)}\hat{\eta}$ and 
$\teq=\tQ^{\l/(\l+1)}\eta_{\rm eq}$.\footnote{Rough estimation of $\hat{\eta}$
and $\eta_{\rm eq}$ are the followings. As $\hat{t}$ is determined by the condition
such as $|\Psi(\hat{t})|=2|\Psi(0)|$, 
$\sqrt{\hat{\eta}^\l}(v(\hat{\eta})\hat{\eta})^2=$constant for the 2D case.
On the other hand, as  $\teq$ is the time at which the overlap of SF bubbles starts
\cite{SKHI}, $v(\eta_{\rm eq})\eta_{\rm eq}=$constant.
Simulation for various $\l$'s is a future work.}
Furthermore, a typical linear size of the bubble at $t$, i.e., the correlation length at $t$, 
$\xi(t)$, is given as
\be
\xi(t)=\int_0^{t} v_t dt\propto {1 \over \tQ^{\l/2}} \ t^{\l/2+1},
\ee
and therefore, $\xi(\hat{t})\propto (\tQ)^{\l/2(\l+1)}\hat{\eta}^{(\l+2)/2}$ and 
$\xi(\teq)\propto (\tQ)^{\l/2(\l+1)}\eta_{\rm eq}^{(\l+2)/2}$.
{\em After $\teq$, the merging and coarsening process of SF bubbles takes 
place \cite{SKHI}, and therefore the above picture and also the resultant
scaling laws do not hold anymore.} 


\subsection{GL theory, GW methods and quantum Monte-Carlo simulation}

Here, it is suitable to comment on the GW approximation.
The GL theory and also the Gross-Pitaevskii (GP) equation consider only
the mean field and totally ignore fluctuations around it.
On the other hand in the GW approximation, we focus on a wave function
of site factorization, and wave function at each site is obtained by 
solving the site-factorized Hamiltonian in which 
the NN operators are replaced with their expectation values \cite{SKHI}.
The uncertainty relation between the particle number and phase at each site is faithfully
taken into account although an equation of motion similar to the GL (GP) equation
is derived by the GW methods.
This is an advantage of the GW approximation over the GL and GP theories.

As more reliable methods, let us consider the quantum Monte-Carlo (MC) simulations
of the coherent-state path integral in the imaginary-time formalism.
In this MC simulations, quantum operators are reduced into classical variables
and the quantum superpositions are treated by the fluctuations in the imaginary-time 
direction.
Large number of configurations are generated by the MC updates and 
physical quantities are calculated by averaging them over generated
configurations.
In the Metropolis MC algorithm, the local updates are applied to variables at each site
by calculation a local energy around that site. 
In the vicinity of a phase transition point, a large number of configurations contribute
equally, and calculations by large CPU times are required in order to take into
account all relevant configurations.
On the other hand away from the critical point, the number of important
configurations is not so large.
From the viewpoint of the MC simulation with the local update, we can get
an interesting insight into the GW approximation.
That is, let us imagine that we perform a GW calculation for a system with size
$10^4\times 10^4$.
When we calculate expectation values, we divide the $10^4\times 10^4$ system
into $10^4$ number of $10^2\times 10^2$ subsystems.
We obtain the expectation values by averaging values calculated 
in each subsystem.
Compared with the path-integral MC simulation, this method is more reliable
as the uncertainty relation is faithfully respected.
[In the path-integral MC simulation, this relates to the problem
how accurately effects of the Berry phase are taken into account.
See for example, Ref.~\cite{KSI}.]
However in the vicinity of the phase transition, $10^4$ configurations are not 
sufficient to obtain physical quantities closely related to the singularities of 
the phase transition.
The above consideration suggests that the GW methods are a fairly good 
approximation for calculating physical quantities that are finite even for 
the critical regime, e.g., finite order parameters.
In other words, the estimation of the critical exponents by the GW methods
is not reliable even for using very large systems.

The above consideration may over estimate the reliability and applicability
of the GW methods, but it explains why the GW methods often succeed
in obtaining correct results such as the phase diagrams, etc.
We expect that the GW methods also works for the correlation functions
as far as the correlation length is finite as the quantum MC simulations do, 
although at present there are no ways to verify it in the quench dynamics.

\subsection{First-order phase transition in vicinity of triple point}

As the phase diagram in Fig.~\ref{groundstate} shows, the present first-order
phase transition is located in the vicinity of the triple point of the DW,
SF and SS.
The GL theory for the quench dynamics in Sec.~4.1 can be applied to this case
with some modification.
Besides the SF order parameter, we introduce a coarse-grained real DW order
parameter, $D(\vec{r},t) [\sim (-)^in_i]$.
GL equations are given as 
\be
&&{\partial \phi \over \partial t}=\nabla^2_r \phi-\epsilon(t)\phi -g_1|\phi|^2\phi
-g_3D^2\phi, \label{GLeq2} \\
&&{\partial D \over \partial t}=\nabla_r^2D+m(t)D-g_2D^3-2g_3D|\phi|^2,
\label{GLeq3}
\ee
where the positive parameters $g_1, g_2$ and $g_3$ are phenomenological ones,
which are to be determined by the parameters $U$ and $V$.
The positivity of $g_3$ comes from the fact that the SF and DW are competing
orders in the original EBHM.
On the other hand, $\epsilon(t)$ and $m(t)$ are parameters that are determined
by $J(t)$, $U$ and $V$.
In the quench from the DW to SF, both $\epsilon(t)$ and $m(t)$ are decreasing
functions of $t$.

Let us consider a slow quench, and denote the phase transition time from
the DW to SF by $t_c$.
At $t=t_c-\delta \ (\delta \to +0)$, the system is in the DW and then,
$\epsilon(t_c)+g_3D^2(t_c)=\epsilon(t_c)+{g_3 \over g_2}m(t_c)>0$, $\phi=0$
and $D^2={m(t_c) \over g_2}$.
On the other hand at  $t=t_c+\delta \ (\delta \to +0)$, the system is in the SF,
and $m(t_c)-2g_3|\phi|^2=m(t_c)+2{g_3 \over g_1}\epsilon(t_c)<0$, 
$|\phi|^2=-{\epsilon(t_c) \over g_1}$ and $D=0$.
From the above equations, we obtain the constraint for the occurrence of
the direct DW to SF
transition such as $2g^2_3>g_1g_2$, and $\epsilon(t_c)<0, \ m(t_c)>0$.
The critical time, $t_c$, is determined by the condition that the potential
energy ${\cal V}=\epsilon(t)|\phi|^2+{g_1 \over 2}|\phi|^4+g_3D^2|\phi|^2
-m(t){D^2 \over 2}+{g_2 \over 4}D^4$ has the same value in the DW and 
SF states at $t=t_c$.
This condition gives $\epsilon^2(t_c)={g_1 \over 2g_2}m^2(t_c)$.
On the other hand,
the triple point is realized by $\epsilon(t_c)=m(t_c)=0$ or  $2g^2_3=g_1g_2$.

Let us focus on the SF for $t\geq t_c$.
In this case, $D=0$ and we only consider the GL equation in Eq.(\ref{GLeq2})
with $D=0$.
We assume the same protocol with Eq.(\ref{epsilont}) and
then, the transformation in Eq.(\ref{scalGL}) can be applied as in the case of
the second-order phase transition.
Correlation length at time $t$ is estimated as 
\be 
\xi(t)=\int_{t_c}^t v_t dt={1 \over \tQ^{\l/2}}(t^{\l/2+1}-t_c^{\l/2+1}).
\label{Lt2}
\ee
The second term on the RHS in Eq.(\ref{Lt2}) comes from the finite jump
of $\phi$ at the critical point and indicates the deviation from the genuine
second-order phase transition.
However for sufficiently small $t_c$ such as $t_c \ll \hat{t}, \ \teq$,
the correlation length satisfies almost the same scaling law with the KZ one.


\section{Dynamics of phase transition from superfluid to density wave}\label{SFtoDW}

This section considers the temporal evolution of the system under a quench 
from the SF to DW. 
We found that behaviors of the system strongly depend on the initial state.
We shall show the results in the following two subsections.

\subsection{Evolution from the GW ground-state of SF}

\begin{figure}[h]
\centering
\begin{center}
\includegraphics[width=9cm]{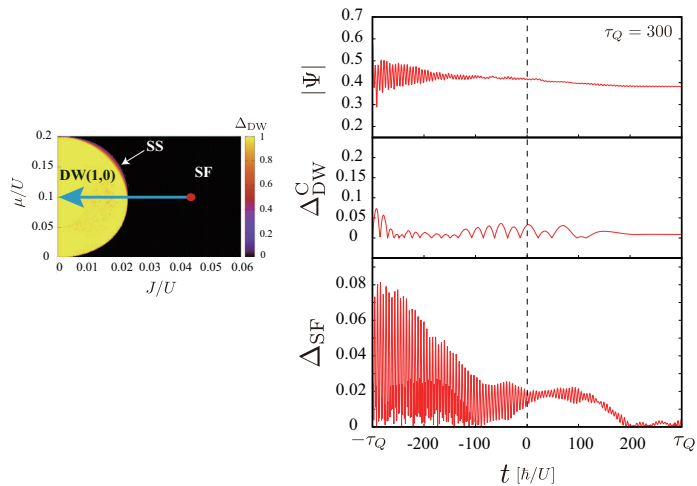}
\end{center}
\caption{Transition from SF to DW with $J(t_{\rm f})=0$, Case A.
The system passes through the critical point $J_c$ at $t=0$.
Even for $t>0$, both the SF amplitude and DW order parameter do not
exhibit the typical behaviors of the DW.
}
\label{SFtoDW1}
\end{figure}

\begin{figure}[h]
\centering
\begin{center}
\includegraphics[width=8.5cm]{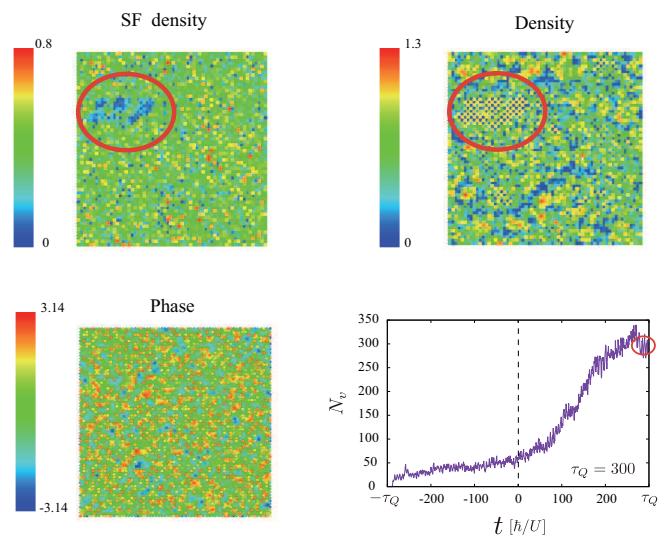}
\end{center}
\caption{Snapshots of SF local density (amplitude), particle density,
SF phase degrees of freedom, and vortex density at $t=\tQ$ ($J/U=0$).
Global coherence of $\Psi_i$ does not exist, and finite-size domains of the DW
partially form as indicated by the red circles.
}
\label{SnapSFtoDW2}
\end{figure}

Let us consider the dynamics of the phase transition from the SF to DW.
The hopping amplitude is varied as follows in the linear quench,
\be
{J_c-J(t) \over J_c}\equiv -\epsilon(t)={t \over \tau_{\rm Q}}.
\label{protocol2}
\ee
In order to clarify the quench dynamics, we shall consider three cases 
in this subsection.
In the first case, Case A, we start with configurations at $J(t=-\tQ)=2J_c=0.044$ 
and terminate the quench at $t=\tQ$ with $J(\tQ)=0$.
We employ the tGW methods to study the system.
In Case A, as well as Cases B and C in the later study in this subsection, 
{\em the initial state is the lowest-energy state obtained by 
the static GW methods}.

The obtained
results of $|\Psi|$, $\Delta_{\rm DW}$ and $\Delta_{\rm SF}$ are shown in 
Fig.~\ref{SFtoDW1} for $\tQ=300$.
$|\Psi|$ exhibits fluctuations in the SF for $t<0$, 
whereas it becomes stable in the region $J<J_c$ (i.e., $t>0)$.
This behavior comes from the fact that $\Psi_i$ has a phase coherence in the SF, 
which induces amplitude fluctuations, as the amplitude and phase of the SF order
parameter are quantum conjugate variables with each other.
On the other hand in the would-be DW region for $t>0$, the phase coherence is lost, 
and then the SF amplitude is stable.
The DW order parameter $\Delta_{\rm DW}$ does not have a stable finite value even 
after passing through the critical point at $t=0$.
These results indicate that some kind of domain structure forms there, i.e.,
small DW domains may coexist with local SF regions.
Calculations of the amplitude of $\Psi_i$ and the particle density at $t=\tQ$
are shown in Fig.~\ref{SnapSFtoDW2}.
As expected above, DW domains and regions with finite SF amplitude coexist
without overlapping with each other.

In Case A, the quench stops with $J(\tQ)=0$, and therefore no movement
of particles occurs after the quench, and the {\em particle-density} snapshot
in Fig.~\ref{SnapSFtoDW2} continues to describe the states for $t>\tQ$.
Similarly, we expect that the coherence of the phase of $\Psi_i$ is
destroyed at $t=\tQ$ because $J(\tQ)=0$ and also $\tQ=300$ is a slow quench.
See Fig.~\ref{SnapSFtoDW2}.
In order to verify the expected behavior of $\Psi_i$, we measured the vortex
density as a function of time. 
At $t=\tQ$, $N_{\rm v}\sim 300$ is sufficiently large.
In summary,
in Case A with $\tQ=300$, an inhomogeneous state with local DW and SF domains
forms after quench.
SF order parameter gradually loses its phase coherence during the slow quench.

On the other hand for cases of smaller $\tQ=100$ and 50, the SF order parameter
$\Psi_i$ is finite even at $t=\tQ$, and it varies after $t=\tQ$.
The {\em phase of $\Psi_i$} gradually loses its long-range coherence by the existence 
of the repulsive interactions for $t>\tQ$.

\begin{figure}[t]
\centering
\begin{center}
\includegraphics[width=8cm]{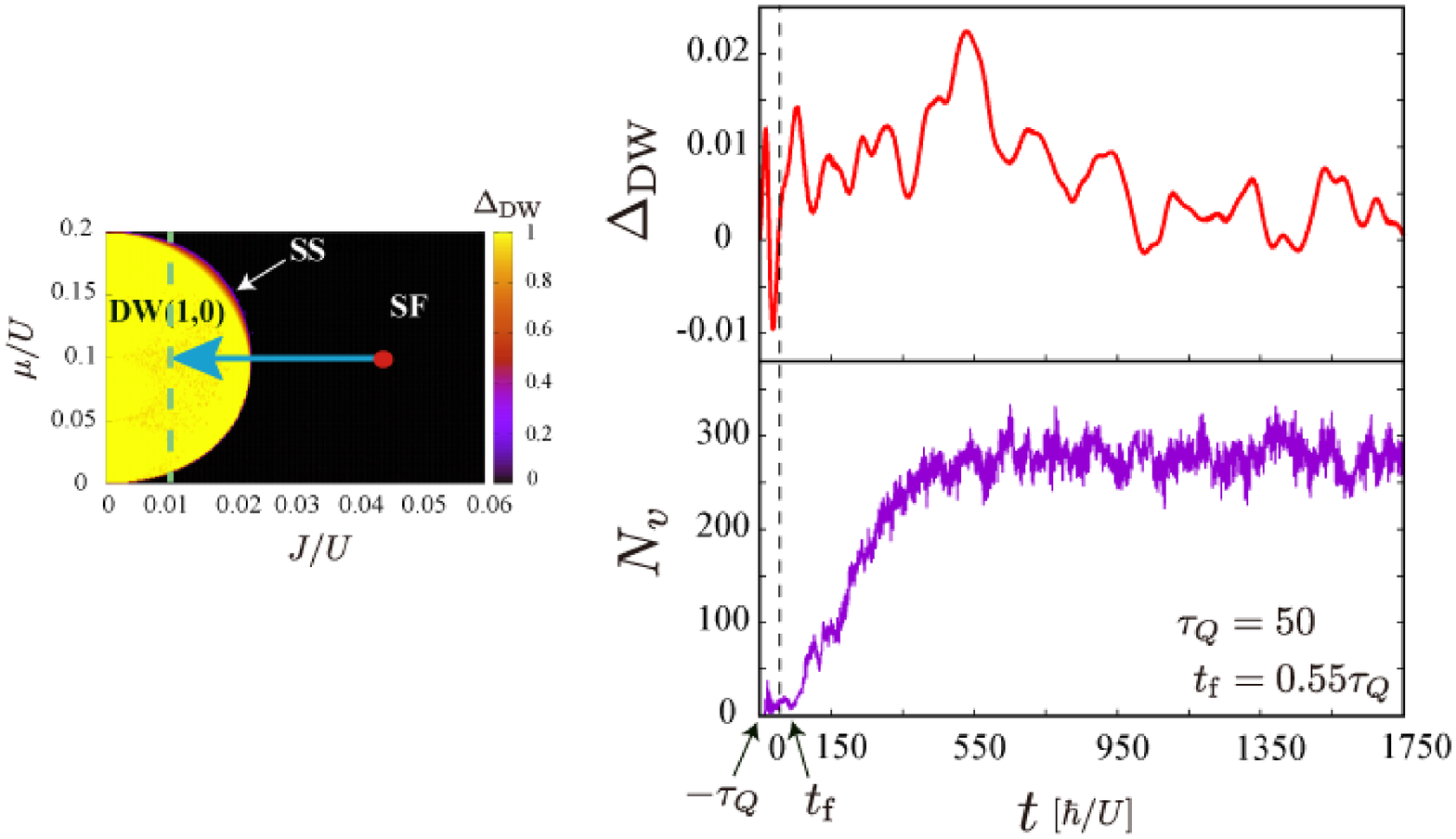}
\includegraphics[width=6cm]{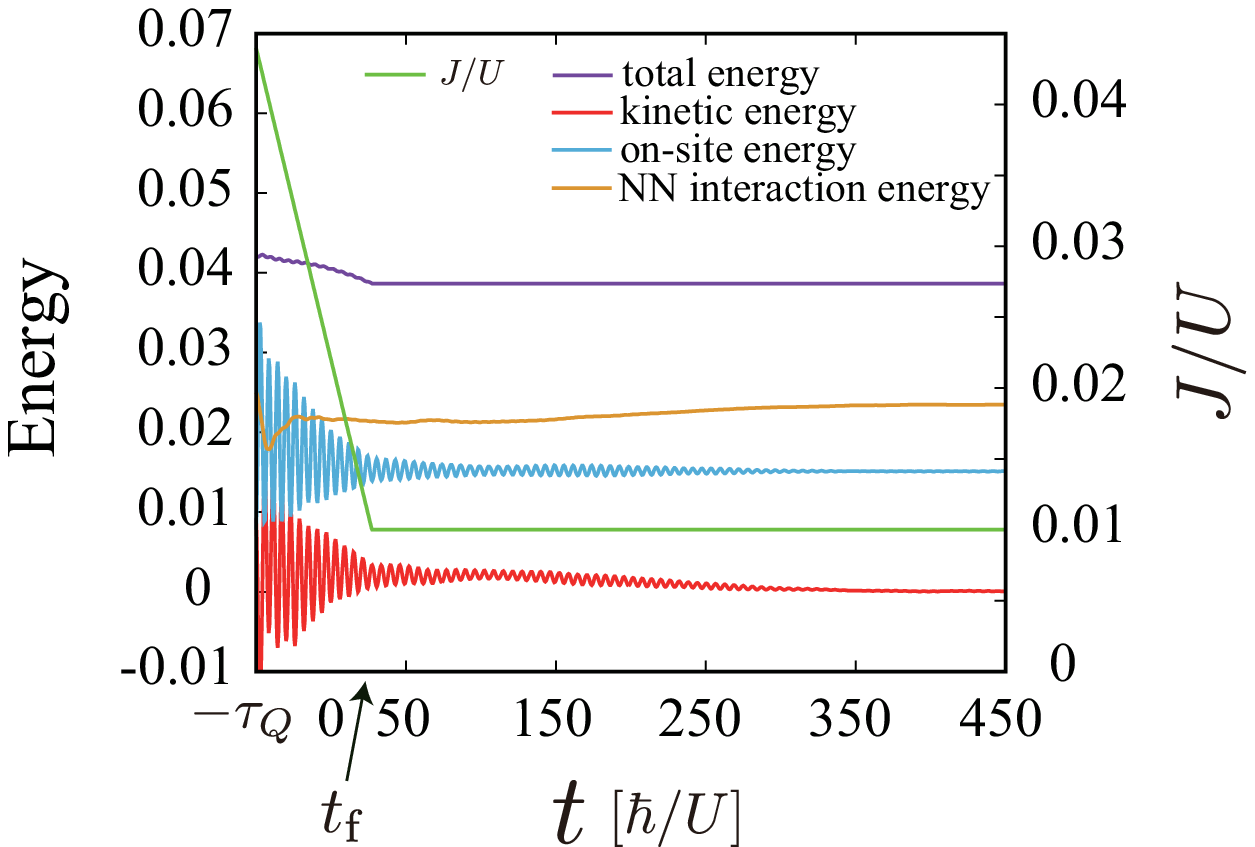}
\end{center}
\caption{Transition from SF to DW with $J(t_{\rm f})=0.01$, Case B.
Genuine global DW order does not form.
After passing $J_c$ at $t=0$, $N_{\rm v}$ keeps a small value for a while, and 
the SF order survives there.
After passing $t_{\rm f}=0.55\tQ=27.5$, the total  energy of the system keeps 
a constant value
as the system is and isolated one.
}
\label{SFtoDW2}
\end{figure}
\begin{figure}[h]
\centering
\begin{center}
\includegraphics[width=8.5cm]{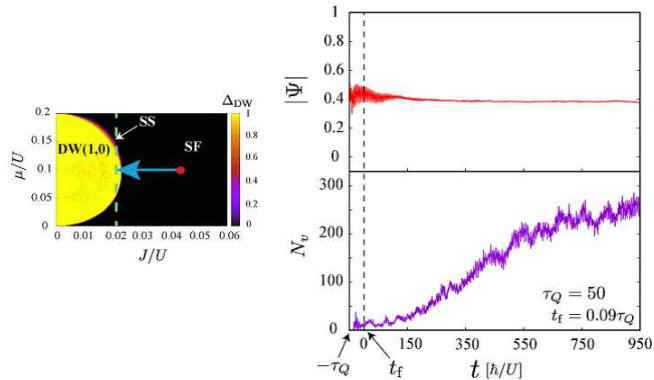}
\end{center}
\caption{Transition from SF to DW with $J(t_{\rm f})=0.02$, Case C.
Increase of $N_{\rm v}$ is slow compared to the cases $J(t_{\rm f})=0$
and $J(t_{\rm f})=0.01$.
SF amplitude $|\Psi|$ also keeps a finite value even for $t\to$large.
However, $N_{\rm v}$ increases smoothly, and therefore, the supercooled
state formed in the quench is not a meta-stable state.
}
\label{SFtoDW3}
\end{figure}
\begin{figure*}[t]
\centering
\begin{center}
\includegraphics[width=6cm]{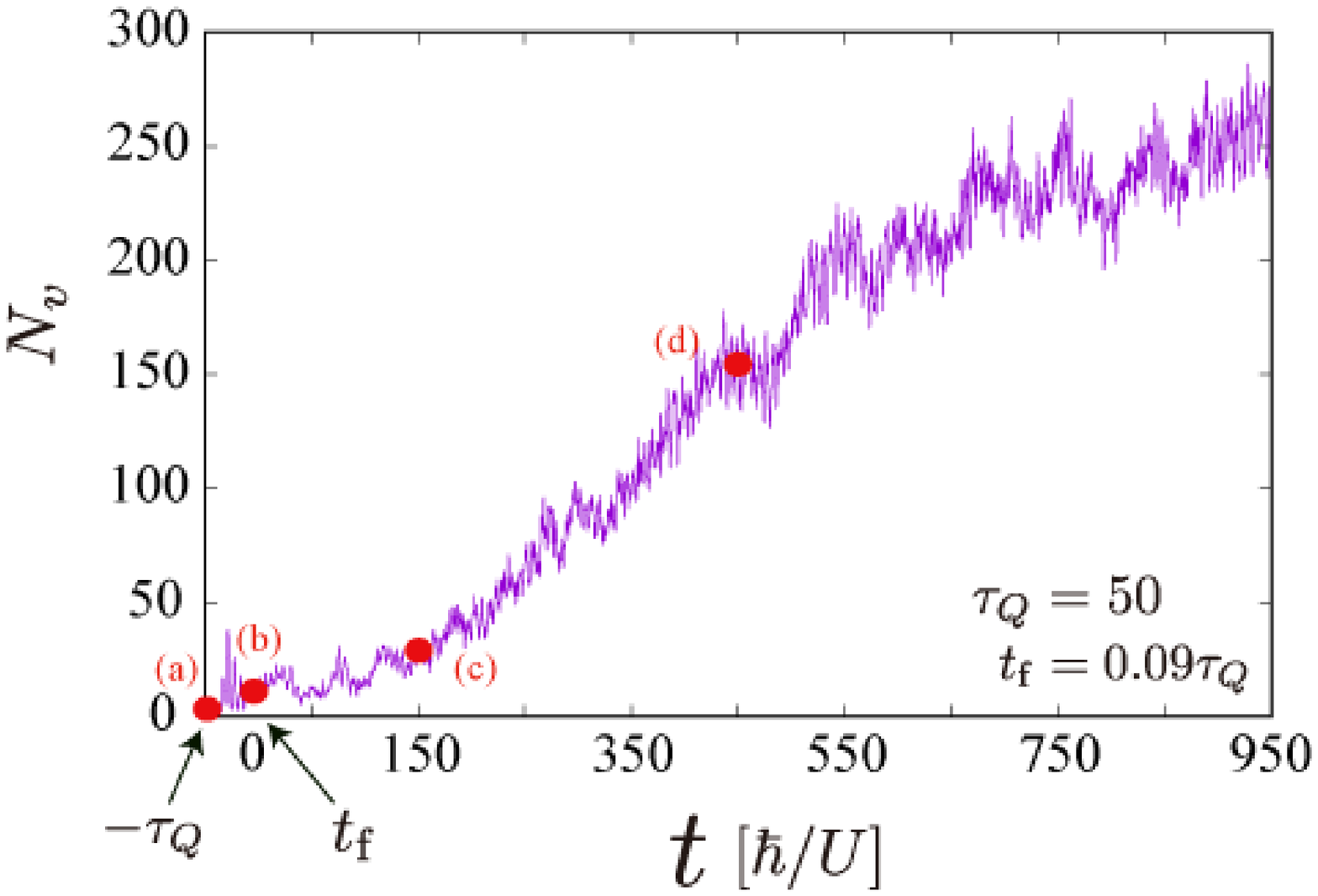} \hspace{0.5cm}
\includegraphics[width=5cm]{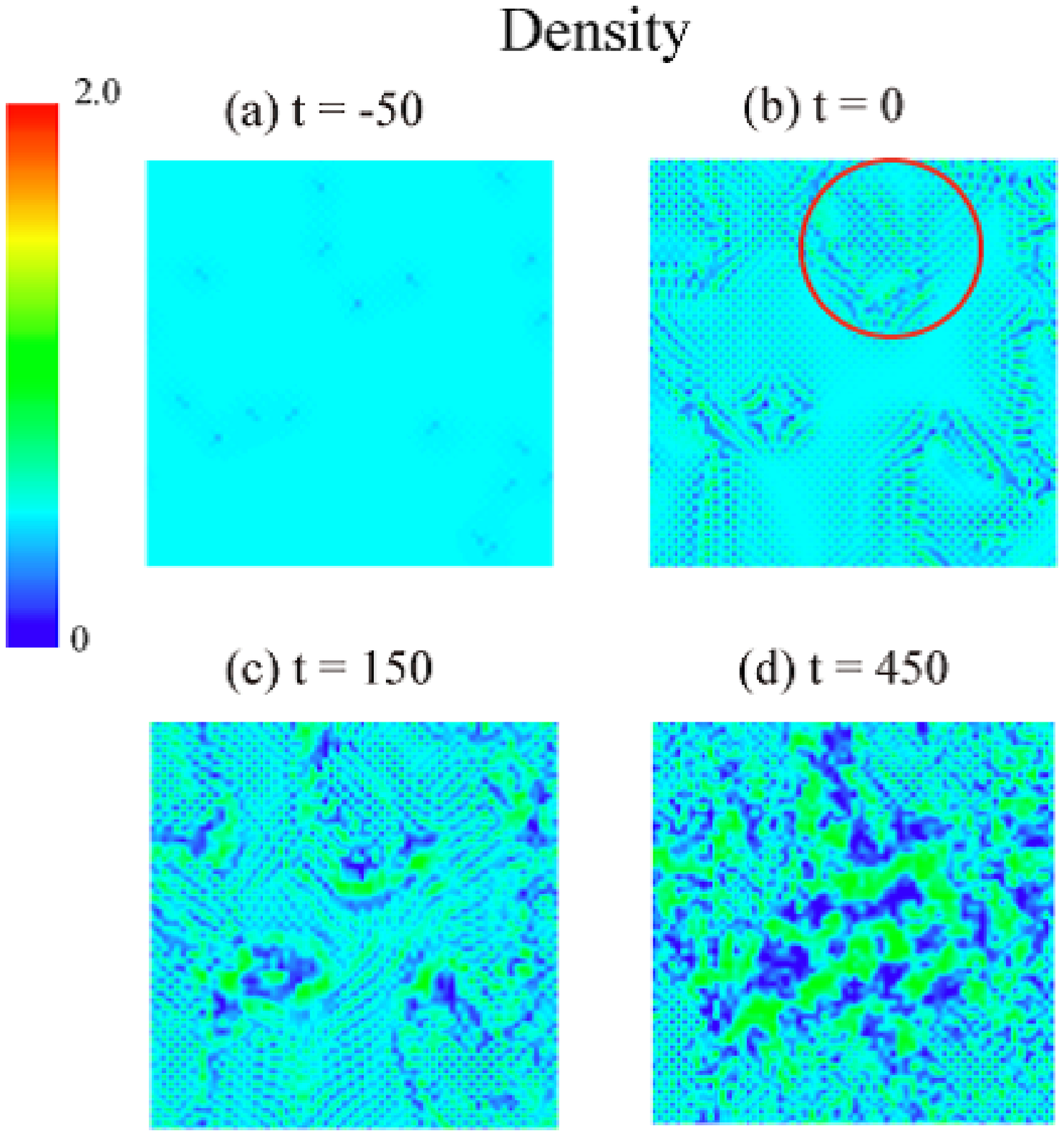} \vspace {1cm} \\
\includegraphics[width=5cm]{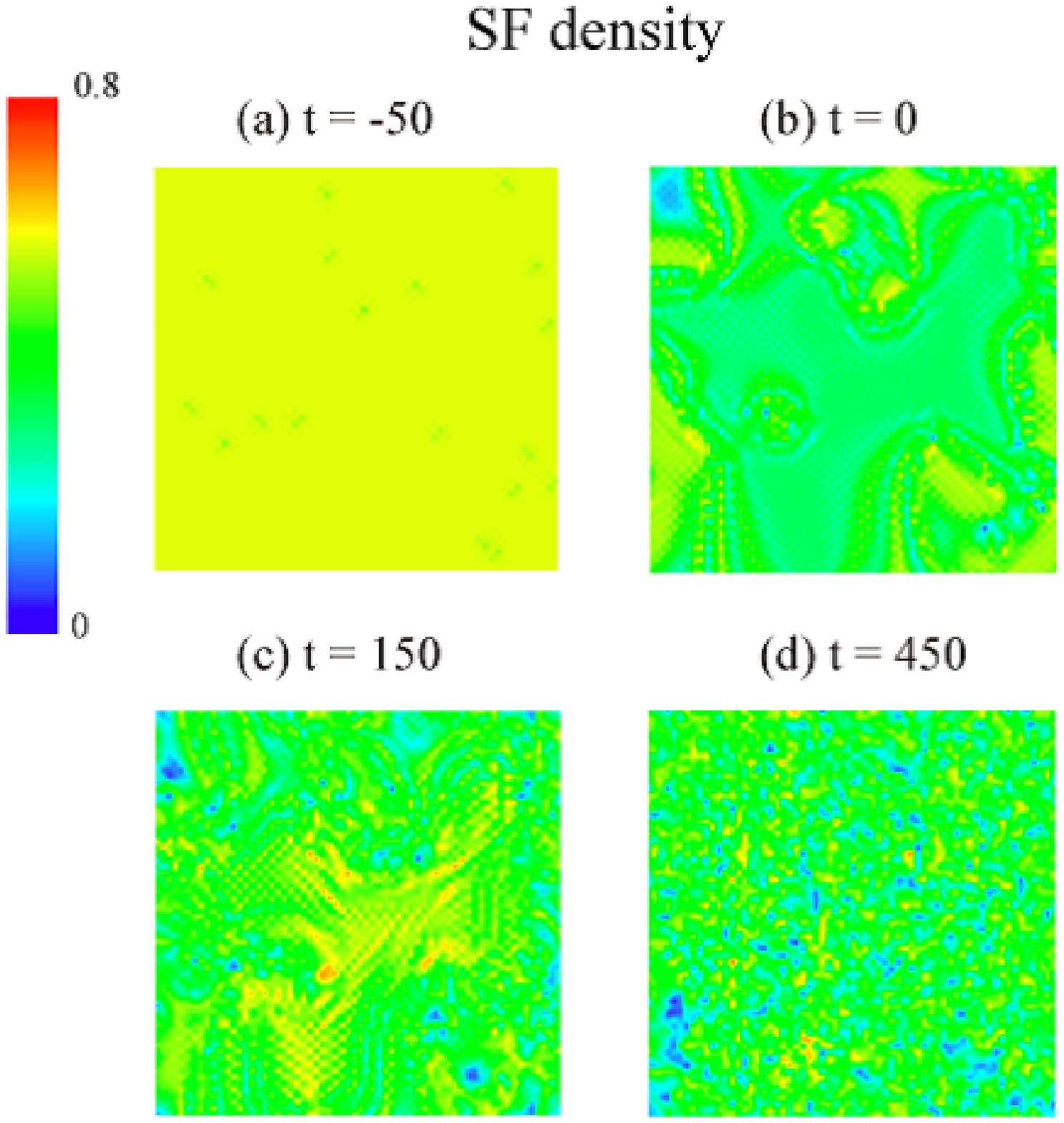}  \hspace{0.6cm}
\includegraphics[width=5cm]{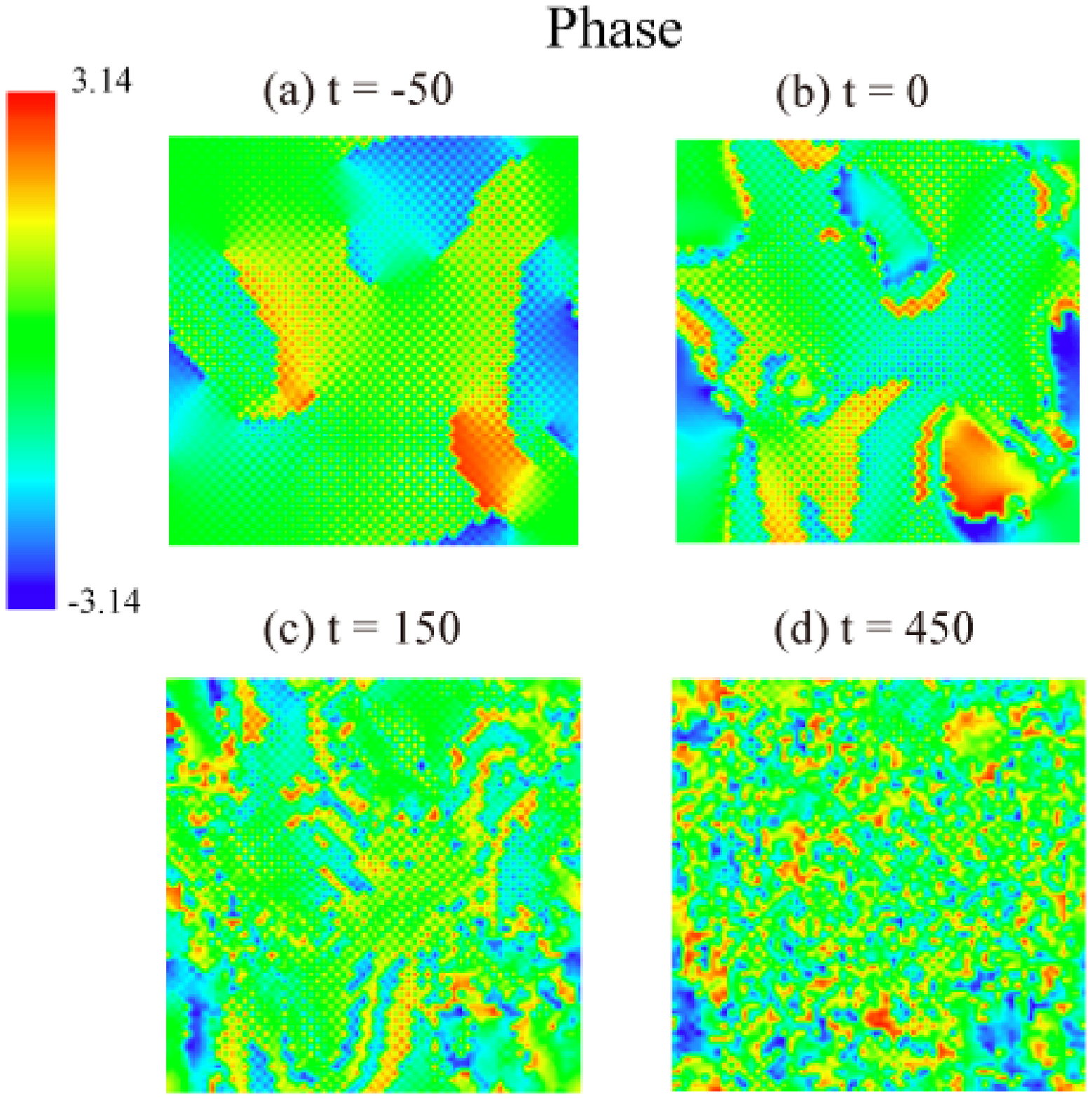}
\end{center}
\caption{(Upper-left) Vortex number as a function of time.
Each point denotes the following time;
(a) $t=-50$, (b) $t=0$, (c) $t=150$, and (d) $t=450$.
(Upper-right) Particle density snapshot in Case C.
At t=0, a typical  DW domain appears as indicated in the red circle. 
(Lower-left) SF density snapshot in Case C.
(Lower-right) Snapshot of phase degrees of freedom of SF order parameter in Case C.
}
\label{CaseCsnap}
\end{figure*}

As Case B, we consider a quench such as $J(-\tQ)=0.044$ and $J(0)=J_c=0.022$
as before but it terminates at $t=t_{\rm f}$ with $J(t_{\rm f})=0.01$, i.e., 
$t_{\rm f}=0.55\tQ$ (see Fig.~\ref{SFtoDW2}).
We also study how the system evolves after $t_{\rm f}$.
Observed quantities are shown in Fig.~\ref{SFtoDW2} for $\tQ=50$.
The DW order parameter $\Delta_{\rm DW}$ develops but its value fluctuates 
in rather long period after passing $J_c$ as in Case A.
The total energy slightly decreases until $t_{\rm f}$, and the kinetic and 
on-site energies exhibit fluctuating behavior for $t<t_{\rm f}$ although 
the NN interaction energy is rather stable.
This behavior mostly originates from the local density fluctuations,
and the stability of the NN interaction comes from the cancellation
mechanism between NN sites $j\in i{\rm NN}$. 
After passing the critical point at $t=0$, the $\Psi_i$ keeps a coherent 
SF order for some period as the calculation of the vortex number $N_{\rm v}$ indicates.
At $t\approx 100$, it starts to lose the coherence and the SF is destroyed
as the increase in $N_{\rm v}$ indicates.
The state at $t\sim t_{\rm f}$ is a {\em supercooled state}, and
a coexisting phase of local domains of the DW and SF is realized there.
The observed phenomenon after $t>t_{\rm f}$, therefore, has very similar nature to
the {\em glass transition}, in which
the phase coherence and superfluidity are getting lost as the supercooled state
evolves after the quench.
We call it {\em quantum glass transition (QGT)} as the hopping amplitude $J$, instead of
temperature, is the controlled physical quantity and the relevant transition
is quantum mechanical one instead of thermal one.
We have verified that similar phenomenon is observed for other values of $\tQ$, 
e.g., $\tQ=20$ and $200$.

In both Case A and Case B, the above mentioned QGT is observed {\em dynamically
as a nonequilibrium phenomenon}, i.e., the QGT point is passed through as the system
evolves.
Therefore as the next problem,
it is interesting to see whether there exits a genuine glass transition point,
$J_{\rm g}(<J_c)$. 
Below $J_{\rm g}$, the supercooled state is meta-stable or at least has a long life time,
and the SF survives without losing its phase coherence.
For Cases A and B, $J<J_{\rm g}$.
Then as Case C, we studied the quench whose finial point is 
$J(t_{\rm f})=0.02$, i.e., very close to the equilibrium critical point.
Obtained order parameter $|\Psi|$ and vortex number $N_{\rm v}$ are shown
in Fig.~\ref{SFtoDW3} for $\tQ=50$, and time evolution of the particle density, 
amplitude and phase of $\Psi_i$ are shown in Fig.~\ref{CaseCsnap}.
After passing the critical point $J=J_c$ at $t=0$, the domain formation of 
the DW starts as shown by the particle-density snapshot in Fig.~\ref{CaseCsnap}, 
whereas the long-range coherence of the SF order parameter
$\Psi_i$ exists there.
Compared with the cases of $J(t_{\rm f})=0$ and $J(t_{\rm f})=0.01$,
the destruction of SF and formation of the DW region are slow, but 
after $t>450$, the quantum glass state forms.
Local DW domains develop but also empty regions (voids) form.
SF order loses a long-range coherence.
This result indicates that $J_{\rm g}$ cannot be observed.
Similar results are obtained for the case of $\tQ=20$ and $\tQ=200$.


\subsection{Evolution from SF state with small phase fluctuations}

\begin{figure*}[t]
\centering
\begin{center}
\includegraphics[width=6cm]{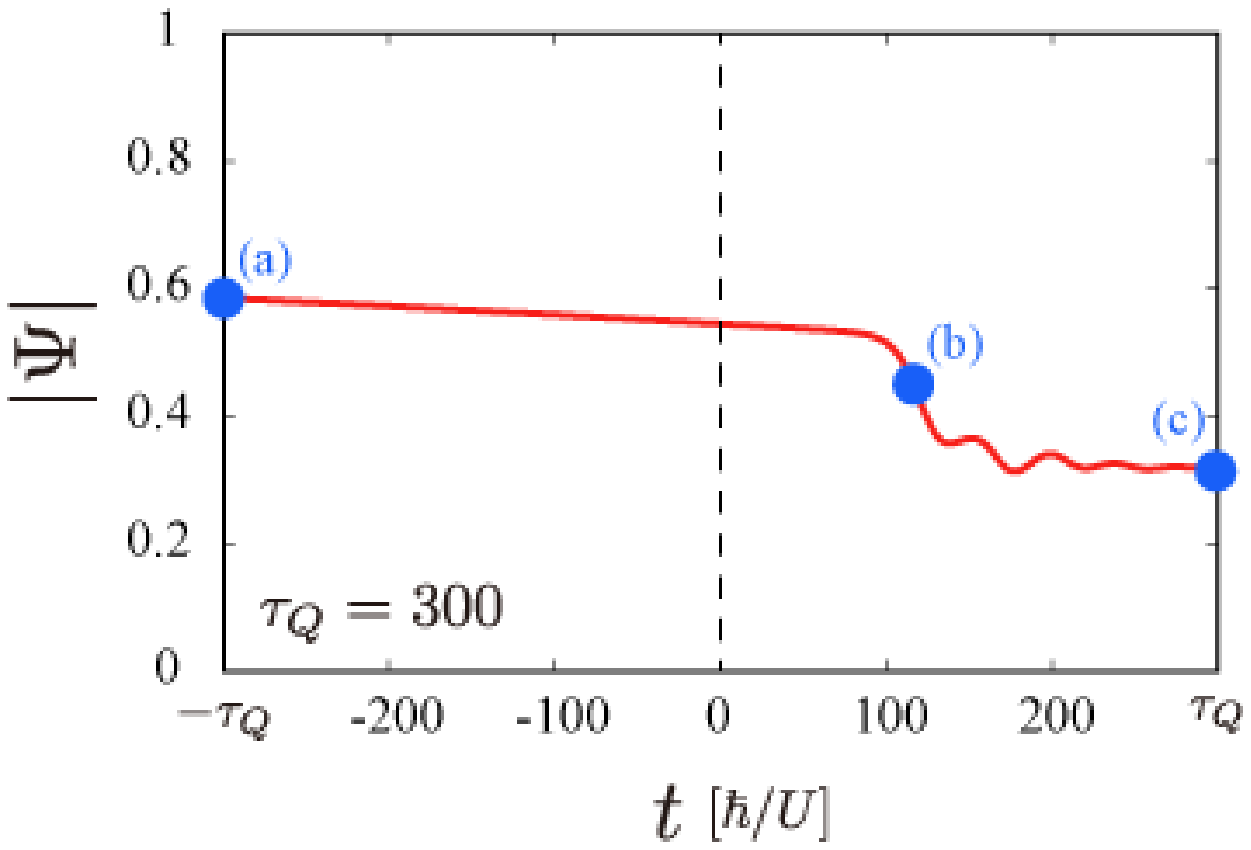}
\includegraphics[width=12cm]{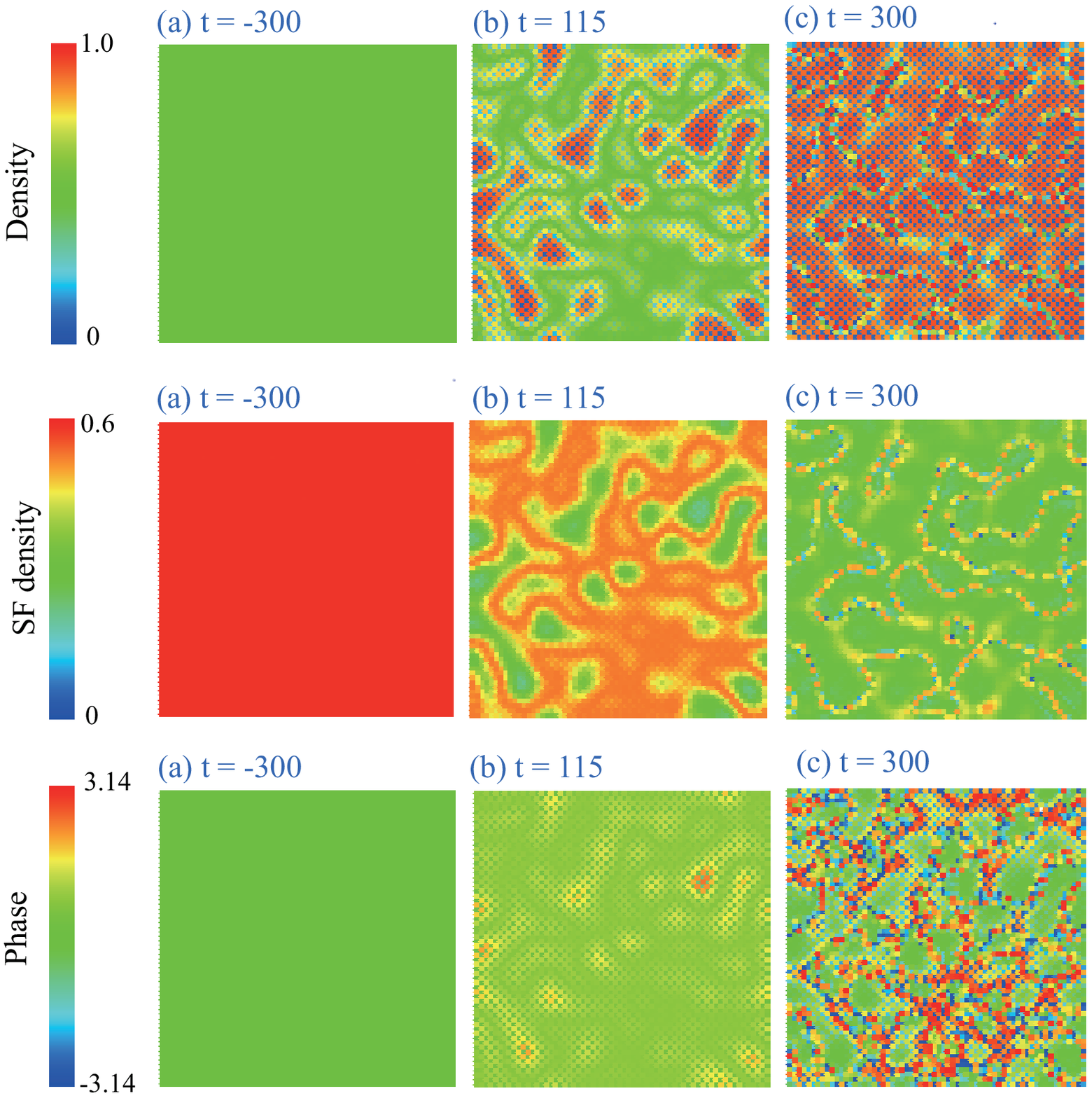}
\end{center}
\caption{(First) SF order parameter as a function of time.
Each point denotes the following time;
(a) $t=-300$, (b) $t=115$, and (c) $t=300$.
(Second) Particle density snapshot in Case D.
(Third) SF density snapshot in Case D.
(Lowest) Snapshot of phase of SF order parameter in Case D.
At $t=300$, a large scale DW domain structure with thin domain walls forms.
Coherence of SF phase is lost there.
}
\label{CaseDsnap}
\end{figure*}

In Sec.~5.1, we studied dynamical evolution of the system from the SF to DW.
In that study, the initial state is set to the ground-state obtained by
the equilibrium GW methods.
It is interesting to see how the dynamical phenomena depend on the initial state
as we are considering the first-order phase transition.
In order to study this problem, we consider a SF state that is uniform and
has almost perfect phase coherence with very small random fluctuations.
For the practical calculation, we employ an initial state GW wave function in Eq.(\ref{GW})
corresponding to $\Psi_j=\sqrt{\rho}e^{i\delta\theta_j}$ 
with random numbers $\{\delta\theta_j\}$ from a uniform distribution
$[-0.005,0.005]\times\pi$.
The other condition is the same with the Case A, (please refer to the left panel 
in Fig.~\ref{SFtoDW1}).
We call the present study Case D.

We investigated the time evolution of the system by the tGW methods, and 
obtained results are shown in Fig.~\ref{CaseDsnap}.
Interestingly enough, the system behavior after passing across the critical point
$J_c$ is substantially different from that in Cases A.
The SF order parameter $|\Psi|$ decreases a finite amount at $t\sim 100$,
and the density difference at even-odd sublattice increases there.
On the other hand, the vortex number starts to increase rapidly at $t\sim 150$.

Snapshots of the particle density, SF amplitude and SF phase are shown in 
Fig.~\ref{CaseDsnap}.
Contrary to Case A, the DW pattern starts to form at $t\sim 115$ and 
it develops to the whole system at $t\sim 300$, even though there
exist domain walls.
It should be noticed that
a similar behavior was observed for the classical first-order phase 
transition in Ref.~\cite{firstPT1}.
On the other hand, the SF phase coherence exists at $t<115$, whereas
it is destroyed at $t\sim 300$.

The initial state of Case D has higher energy than that of Case A.
The above numerical result indicates that there exists an energy barrier
between the supercooled SF state and the genuine DW, and some amount
of energy is need to overcome the barrier.
Furthermore, the above result also indicates that the existence of the SF
phase coherence in large spatial regions prevents the formation of large size DW domains. 
In other words, local fluctuations of the superfluidity coherence substantially develops 
under a quench even if they are initially tiny, and the DW is preferred as a result.

We expect that the above interesting phenomenon is observed by experiments on 
ultra-cold atomic gases in the near future. 


\section{Conclusion}\label{conclusion}

In this work, we studied dynamical behavior of the EBHM in 2D by using
the tGW methods.
In the ground-state phase diagram, there are three phases, the SF, DW, and SS.
In particular, we are interested in the first-order phase transition between
the SF and DW under a slow quench of the hopping amplitude.

First, we investigated the dynamics of the EBHM in the transition from the DW to SF.
In the practical calculation, we fix the strength of the one-site and NN repulsions,
and vary the hopping parameter $J$.
After passing through the equilibrium critical point $J_c$, the amplitude of
the SF order parameter, $|\Psi|$, remains vanishingly small until $t=\hat{t}$.
After $\hat{t}$, it develops quite rapidly.
Therefore, $\hat{t}$ has the meaning of the reentry time to the adiabatic region
passing from the frozen regime although the present phase transition is of first order.
At $\teq(>\hat{t})$, $|\Psi|$ stars to oscillate until $t=t_{\rm ex}$.
This behavior is quite similar to that in the second-order phase 
transition from the Mott insulator to SF, which we observed in the previous 
work~\cite{SKHI}.
Then we are interested in whether some kind of scaling laws between 
the correlation length/vortex number and the quench time $\tQ$ exist.
Our numerical study shows that the scaling laws such as $\xi\propto \tQ^b$
and $N_{\rm v}\propto \tQ^{-d}$ in fact hold.
This result is against to the simple expectation that such scaling laws do not exist
in the first-order phase transitions because the simple relaxation-time picture
and the concept of the (dynamical) critical exponents are not applicable.
From this result, we think that there exists another mechanism, besides the KZ
mechanism, to generate the scaling laws.
As a possible explanation, we studied the present system by using the GL-type theory
suggested by Ref.~\cite{Niko}.
This consideration indicates that the observed scaling laws come from the fact
that the present phase transition point is located in the vicinity of the triple point.

In the second half, we studied the dynamics of the EBHM in the quench of
the opposite direction, i.e., from the SF to DW.
We focused on how the final value of the hopping amplitude of the quench,
$J(t_{\rm f})$, influences the dynamics of the system during and after the quench.

Our numerical study showed very interesting phenomena.
First, in the case for the GW ground-state as the initial state, the
genuine DW state does not form even for very slow quench $\tQ=300$.
Instead, the coexisting state composed of DW and SF domains appears and 
spatially inhomogeneous structure of that state is stable after the quench.
In cases with $J(t_{\rm f})>0$, the SF order parameter has a phase coherence
at $t=t_{\rm f}$, and after the quench, the SF order is getting weak by the 
generation of vortices. 
Obviously, the quench produces a {\em supercooled state} in which the domain
structure of the DW and SF local (i.e., short-range) coherent state forms.
These two domains have an off-set structure with each other.
Then, after termination of the quench, the SF is destroyed.
This phenomenon is a reminiscent of the glass transition in classical
polymers etc, and we call the observed phenomenon quantum glass transition.

On the other hand, if we start with the uniform SF state with tiny fluctuations in the 
phase of the SF order, the system evolves into the DW with thin domain walls.

In the phase diagram of the EBHM near the half-filling shown in 
Fig.~\ref{groundstate}, there is the SS phase, and the SS has two
phase boundaries with the DW and SF.
In the case of the mean particle density $\rho=1$ and strong NN repulsion,
the region of the SS is large and two second-order phase transitions are 
observed clearly from the SS to the DW and SF, respectively.
It is interesting to study the dynamics in that region, that is, how the system
develops crossing through two second-order phase boundaries.
Some related problem was recently studied in classical systems, and 
a modified KZ scaling law was proposed~\cite{hybrid}.
We studied the above problem in the EBHM by using tGW methods, 
and results are published in Ref.~\cite{KZIII}.


\appendix
\renewcommand{\thefigure}{A.\arabic{figure}}
\setcounter{figure}{0}
\renewcommand{\theequation}{A.\arabic{equation}}

\section*{Appendix A. Hard-core Bose-Hubbard model}

\begin{figure}[h]
\centering
\begin{center}
\includegraphics[width=6cm]{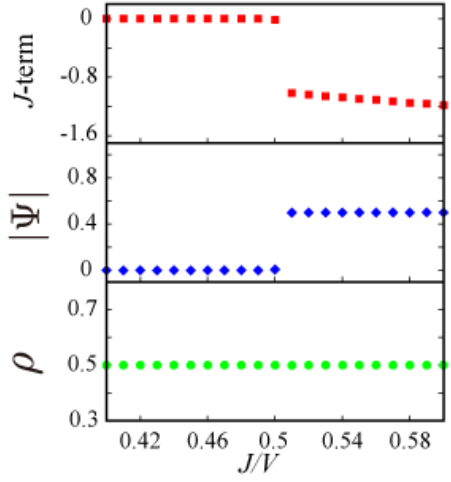}
\end{center}
\caption{Equilibrium physical quantities obtained by the GW methods.
$\rho=1/2$ and $V=1$. The results show the existence of a first-order
phase transition as the quantum simulations in Refs.~\cite{HC1,HC2} proved.
}
\label{HCcal}
\end{figure}
\begin{figure}[h]
\centering
\begin{center}
\includegraphics[width=6cm]{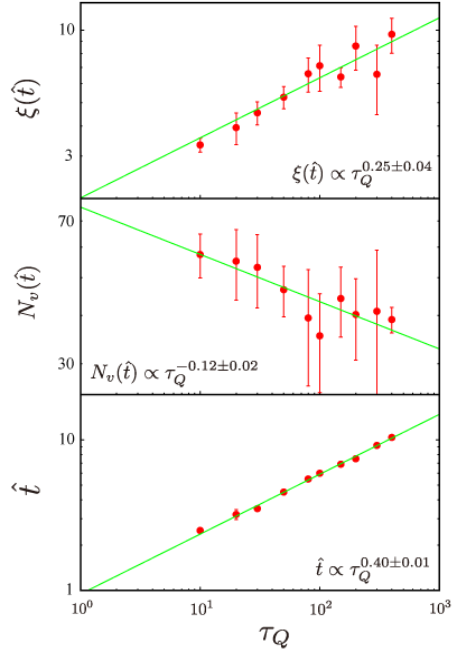}
\end{center}
\caption{Quench dynamics of the HCEBHM with
$\rho=1/2$ and $\tQ=300$. 
The correlation length $\xi$ and vortex density $N_{\rm v}$ fluctuate
rather strongly compared to the soft-core cases.
This result comes from the fact that particle-number fluctuation at each site
is restricted by the hard-core constraint, and as a result, fluctuation in the phase
of the SF order parameter $\Psi_i$ is getting large.
}
\label{HCtQ}
\end{figure}

In this work, we study the EBHM of the soft-core boson.
Hard-core extended Bose-Hubbard model (HCEBHM) is also an interesting model and 
its relationship to the $s=1/2$ quantum spin model is often discussed.
Hamiltonian of the HCEBHM on the square lattice is given as 
\begin{eqnarray}
H_{\rm HC}&=&-J\sum_{\langle i,j \rangle}(a^\dagger_i a_j+\mbox{H.c.})
+V\sum_{\langle i,j \rangle}n_in_j-\mu\sum_in_i,
\label{HCBH}
\end{eqnarray}
where the on-site interaction terms do not exist by the hard-core nature.
Phase diagram of the model in Eq.(\ref{HCBH}) was studied by the quantum
MC simulations \cite{HC1,HC2}, and it was verified that a first-order phase transition
between the DW and SF exists at half filling $\rho=1/2$ as in the soft-core case. 
Then, it is interesting to study the quench dynamics of the HCEBHM by the GW
methods.

In this appendix, we shall give numerical calculations of the physical quantities
concerning to the static properties of the model in Fig.~\ref{HCcal}, and also 
the $\tQ$-dependence of $\hat{t}$, etc in the quench dynamics in Fig.~\ref{HCtQ}.
The results in Fig.~\ref{HCcal} obviously show that there is a first-order
phase transition from the DW to SF for increasing $J/V$ as the quantum MC simulations 
in Ref.~\cite{HC1,HC2} proved.
On the other hand,
the correlation length $\xi$ and vortex density $N_{\rm v}$ fluctuate
rather strongly compared to the soft-core cases.
This result comes from the fact that the HCEBHM has a small fluctuations in the 
particle number at each site, and as a result, the phase of the SF order parameter
fluctuates rather randomly.


\section*{Acknowledgments}
Y. K. acknowledges the support of a Grant-in-Aid for JSPS
Fellows (No.17J00486).


\end{document}